\documentclass[11pt,a4paper]{article}
\usepackage{amstext,amsmath,amsfonts,amsbsy,amssymb,amscd,bbm,epsfig,sint}

\newcommand{\rmP}{{\rm P}}
\newcommand{\rmS}{{\rm S}}

\newcommand{\unit}{1\kern-.25em {\rm l}}

\newcommand{\be}{\begin{equation}}
\newcommand{\ee}{\end{equation}}
\newcommand{\bd}{\begin{displaymath}}
\newcommand{\ed}{\end{displaymath}}
\newcommand{\bea}{\begin{eqnarray}}
\newcommand{\eea}{\end{eqnarray}}
\newcommand{\ba}{\begin{array}}
\newcommand{\ea}{\end{array}}
\newcommand{\req}[1]{eq.~(\ref{#1})}
\newcommand{\res}[1]{section~\ref{#1}}
\newcommand{\rep}[1]{\cite{#1}}

\newcommand{\Tr}{{\rm Tr}}
\newcommand{\tr}{{\rm tr}}

\newcommand{\psibar}{\bar{\psi}}

\newcommand{\Real}{\relax{\mathsf{\Gamma\kern-.35em R}}}
\newcommand{\Int}{\relax{\mathsf{Z\kern-.40em Z}}}

\newcommand{\ci}[1]{\boldsymbol{#1}}

\newcommand{\smm}{\mathbf{m}}
\newcommand{\tmm}{\ci{\mu}}
\newcommand{\diag}{{\rm diag}}
\newcommand{\taa}{\alpha}
\newcommand{\tab}{\beta}

\newcommand{\tata}{\tilde{\taa}}
\newcommand{\tatb}{\tilde{\tab}}

\newcommand{\cA}{{\cal A}}

\newcommand{\cC}{{\cal C}}

\newcommand{\cH}{{\cal H}}

\newcommand{\cP}{{\cal P}}

\newcommand{\Nf}{N_{\rm f}}
\newcommand{\MeV}{{\rm MeV}}

\newcommand{\bfp}{{\bf p}}
\newcommand{\bfq}{{\bf q}}
\newcommand{\bfx}{{\bf x}}

\newcommand{\bfz}{{\bf z}}

\newcommand{\rme}{{\rm e}}
\newcommand{\rmO}{{\rm O}}

\newcommand{\rmR}{{\rm R}}

\begin{document}
\begin{titlepage}
\begin{flushright}
   DESY 04-079\\
   FTUAM-03-22\\
   IFT UAM-CSIC/03-31\\
   ROM2F/2004/10\\
\end{flushright}
\vskip 7ex
\begin{center}
  {\Large\bf Twisted mass QCD and lattice approaches\\[1.5ex]
   to the  $\Delta I = 1/2$ rule}
\end{center}
\vskip 5ex
\begin{center}
{\large Carlos Pena$^{\scriptscriptstyle a}$,
        Stefan Sint$^{\scriptscriptstyle b,c}$
  and Anastassios Vladikas$^{\scriptscriptstyle d}$}
\end{center}
\vskip 3ex
\begin{flushleft}
$^{\scriptstyle a}$ DESY, Theory Group, Notkestra{\ss}e 85, 
D-22603 Hamburg, Germany\\[1ex]
$^{\scriptstyle b}$ Departamento de F\'{\i}sica Te\'orica,
Facultad de Ciencias C-XI,\\
$\hphantom{^{\scriptstyle a}}$  Universidad Aut\'onoma de Madrid,
E-28049 Cantoblanco, Madrid, Spain\\[1ex]
$^{\scriptstyle c}$ Instituto de F\'{\i}sica Te\'orica, CSIC/UAM,
Facultad de Ciencias C-XVI,\\
$\hphantom{^{\scriptstyle a}}$ Universidad Aut\'onoma de Madrid, E-28049 Cantoblanco, Madrid, Spain\\[1ex]
$^{\scriptstyle d}$ INFN - Sezione di Roma 2,
c/o Dipartimento di Fisica,\\
$\hphantom{^{\scriptstyle a}}$
Universit\`a di Roma ``Tor Vergata'', Via della Ricerca Scientifica 1,\\
$\hphantom{^{\scriptstyle a}}$
I-00133 Rome, Italy\\[5ex]
\end{flushleft}
\vskip 4ex
\begin{center} 
 {\bf Abstract}
\end{center}
\vskip 0.7ex
Twisted mass lattice QCD (tmQCD), generalised to four Wilson quark flavours, 
can be used for the computation of some weak matrix elements related 
to $\Delta I=1/2$ transitions. Besides eliminating unphysical zero modes, 
tmQCD may alleviate the four-quark operator renormalisation problems
encountered in the calculation of CP-con\-serving $K\to\pi$ 
matrix elements with traditional Wilson fermion regularisation.
With an active charm quark, the renormalisation of the
$K\to\pi$ matrix elements requires at most the subtraction of
a linearly divergent counterterm. Furthermore,
in the (partially) quenched approximation the twist angles can be chosen 
so that only a finite counterterm needs to be subtracted.
\vskip 5ex
\begin{center}
    May 2004
\end{center}
\eject

\end{titlepage}

\section{Introduction}
\label{sec:intro}

In the phenomenology of non-leptonic kaon and hyperon decays, the observed 
enhancement of decays with isospin change $\Delta I=1/2$
over those with $\Delta I=3/2$ is referred to
as the $\Delta I=1/2$ rule.
For example, $K^0$ decays into a pion pair of
definite isospin $I$ can be parameterised through
\be
   \cA\left(K^0\to\left(\pi\pi\right)_{I}\right)=A_I{\rm e}^{i\delta_I},
\ee
where $\delta_I$ are the $S$-wave $\pi\pi$ phase shifts.
The $\Delta I=1/2$ rule manifests itself in the large value of
the ratio
\be
   \label{Delta_I_half_rule}
    \vert A_0/A_2\vert \approx 22.
\ee
Although the exact origin of this enhancement is unclear, 
it is believed that it must be related to long-distance, 
non-perturbative physics. The contribution coming from energy 
scales where perturbation theory can be reliably applied 
supplies only a small fraction of this 
ratio\footnote{The precise value depends on the renormalisation
  scheme used in the computation; e.g. in typical $\overline{\rm MS}$ schemes
  it roughly amounts to a factor of two.}. The rest should be an effect
characteristic of low-energy QCD scales.
Analytic studies have provided only a partial understanding 
of this non-perturbative  enhancement factor~\rep{analytic}. 
It is fair to say that we are still far from a 
quantitative prediction based on QCD.

In principle, numerical simulations of 
lattice QCD are an ideal tool for the computation of the required 
hadronic matrix elements. Several lattice regularisations of the
fermionic action has been used to this aim. We cite indicatively
recent work performed with Wilson~\cite{wil}, staggered~\cite{stag},
and domain wall~\cite{GW} fermions; for a more complete and fairly
updated list of references, see ref.~\cite{becir}. In spite of the
considerable effort invested in these simulations and several
encouraging results, the ultimate goal
is still far from being achieved, also because
a number of conceptual and technical problems must be addressed
beforehand.

First, the finite spatial volume, characteristic of lattice simulations,
strongly affects the two-pion final state~\cite{MaianiTesta}.
In spite of this, it has been shown in~\cite{Lellouch:2000pv} that
physical $K \rightarrow \pi \pi$ matrix elements can nevertheless
be obtained from Euclidean correlation functions, computed in a finite volume
(see also \cite{Lin:2001ek} for a related approach).
In particular a relation between the absolute values of the
finite and infinite volume amplitudes is established. This,
together with a separate measurement of the two-pion 
energies in finite volume, allows to fully reconstruct 
the desired decay amplitudes. 
However, the required spatial sizes $L$ still tend to be larger 
by a factor of 2-3 than the volumes typically used for 
the numerical determination of the hadronic
spectrum.\footnote{For a recent proposal, mitigating this requirement,
see ref.~\cite{KimChrist}.}
Furthermore, unitarity is essential for the derivation of these results,
so that the quenched approximation may not be applicable. Therefore,
the practical feasibility of this programme hinges on possible
theoretical developments, as well as further progress in 
simulation algorithms and computer hardware.

The difficulty of computing the physical amplitudes directly
has led to less ambitious approaches, most of which
aim at the determination of the coupling constants in the 
effective chiral Lagrangian~\cite{Bernard:wf}. 
This may be achieved by studying unphysical processes
where no problem with final state 
interactions arises, such as $K\to\pi$ transitions, 
or $K\to\pi\pi$ decays with both pions at 
rest~\cite{Bernard:wf,Bernard:1987pr,dih_review_2}.
Once the effective coupling constants are known, the 
physical amplitudes can be inferred, up to higher order corrections
in the chiral expansion. However, one might expect 
the next-to-leading order corrections
to be large, especially if the matching to data from numerical simulations 
is performed at unphysically large meson masses. 
Assuming that the chiral expansion remains valid in this regime, 
it is thus desirable to include higher order terms in the chiral 
Lagrangian~\cite{Laiho:2002jq,Lin:2002nq}.
Unfortunately, since this introduces many more unknown parameters
already at next-to-leading order, the whole approach becomes 
quite cumbersome if not impractical.

Whatever the chosen strategy, a difficult problem consists in
the renormalisation of the four-quark operators which appear
in the effective weak Hamiltonian.
The details strongly depend on the symmetries of the lattice regularisation; 
regularisations which preserve chiral symmetry are clearly 
advantageous~\cite{Capitani:2000da,GW_theor,Aoki:2000ee}, 
and are increasingly used despite 
their high computational costs. 
If computationally cheaper quarks of the Wilson type are 
used instead, the operator renormalisation 
is complicated, mainly due to the presence of 
power divergences \cite{Bochicchio:1985xa,Maiani:1986db}.
However, it should be noted that power divergences seem
unavoidable, even with chirally symmetric regularisations, if
the charm quark is not included as a dynamical degree 
of freedom in the low-energy approximation to the Standard Model. 

In this paper we advocate the traditional approach of computing
$K\to\pi$ transitions, in a regime where lowest order chiral
perturbation theory can be expected to work reliably. For chirally
symmetric regularisations this requirement seems prohibitive,
even in the quenched approximation, as the volume needs to be 
relatively large\footnote{See, however,
ref.~\cite{Hernandez:2002ds,Giusti:2004an} for an interesting
alternative approach, where the effective theory is 
matched in a finite volume very close to the chiral limit.}.
For Wilson-type fermions this problem is significantly less severe,
but the renormalisation of the four-quark operators is very complicated. 
Even more importantly, in quenched simulations one is again 
forced to use relatively large meson masses, 
due to the presence of unphysical fermion zero modes.
It is the purpose of the present paper to show that both 
the zero mode problem and the problem of operator renormalisation 
can be solved by using twisted mass QCD with four flavours of
Wilson-type quarks. This could provide an affordable 
way of approaching the chiral regime by ``brute force'', 
down to meson masses which are roughly twice the physical pion mass. 
In this regime the chiral expansion can be applied with confidence.

Since the arguments leading to our results are fairly complex,
we have strived to present them in a detailed and fairly
self-contained way. Moreover, in the conclusions we have summarised our
findings in the form of recipes which can be applied in future
numerical computations. Here we only wish to stress the underlying
general principles which are at play behind the simplification
of the renormalisation properties of the $K \to \pi$ matrix elements 
in tmQCD. In this lattice regularisation, a twisted mass term is
introduced in the Wilson fermion action. The ratio between the new
(twisted) and the standard mass terms, expressed by ``twist angles'',
needs to be constrained
in order to recover continuum QCD from the tmQCD regularisation. 
We are free to fix the twist angles to any value, but once the
choice is made it is binding, in that it effectively labels
a specific mode in which chirality is broken by the Wilson term.
This in turn implies that a given operator is in general expressed
in the twisted theory by a combination of all the components
of its chiral multiplet. Since chiral symmetry is broken in the bare
theory, each of these components has different renormalisation
properties. Thus, judicious choices of the twist angles
allow us to map the original operator into one of its chiral partners
which has better renormalisation properties. This is exactly what happens
in the case of the operators characteristic of $K \to \pi$ transitions.
We are able to map the original operator, which in the standard Wilson
fermion regularisation has a quadratic divergence, into one which only
has linear divergences. Since quadratic divergences are known to be
prohibitively hard to subtract in non perturbative computations, while
linear ones are well within our control, this is a significant gain.
This softening of the divergence comes at a modest price: the twisted
theory is characterised by soft symmetry breaking; in particular parity
and isospin (in two directions of isospin space) are lost.
Although these symmetries are recovered in the continuum
limit, their breaking in the bare theory allows for more counterterms
to come into play. In the case under consideration, the breaking
of parity implies that one extra linearly divergent counterterm needs
to be subtracted; i.e. an extra renormalisation condition must be imposed.
This condition amounts to the restoration of parity as the continuum
limit is taken. These considerations are valid in general. 
In more specific cases (choice of twist angles, improved action, 
quenched strange and charm quarks) we have shown that it is possible to 
avoid the linear subtraction altogether,
by working with a pion source at large-time separations. 

The layout of this paper is as follows. After a brief review of the
effective weak Hamiltonian of the Standard Model, and the
effective chiral theory at low energies (sect.~2), we recall
the renormalisation properties of the relevant four-quark
operators both in a chirally invariant regularisation and 
with standard Wilson quarks. In sect.~3 we introduce
twisted mass QCD (tmQCD) with four quark flavours and establish 
the relations between its correlation functions and those of 
standard QCD. Renormalisation and practical aspects of 
lattice regularised tmQCD are discussed in sect.~4.
We are then prepared to describe strategies of how to extract
$K\to\pi$ matrix elements (sect.~5) and we end with
a short summary of our findings (sect.~6). This work has appeared in
preliminary form in refs.~\cite{dih_proc}.

\section{The $\Delta I =1/2$ rule and lattice QCD}
\label{sec:review}

In order to define the general framework we shortly review 
some aspects of the $\Delta S=1$ effective weak Hamiltonian of 
the Standard Model, and its chiral effective theory in
terms of kaons and pions.

\subsection{Effective weak Hamiltonian}

At scales well below $M_W$, but above the charm quark mass, 
weak, CP-conserving, $\Delta S=1$ interactions can
be described with an effective Hamiltonian of the
form\footnote{We follow the standard convention 
and include effective quark bilinear operators 
as counterterms in the renormalised four-quark operators.}
\bea
 \label{H_eff}
 \cH_{\rm eff} &=& V_{ud}V^*_{us}\frac{G_F}{\sqrt{2}}
  \left[C_+(\mu)O_{\rm R}^+(\mu) + C_-(\mu)O_{\rm R}^-(\mu)\right]
  +{\rm h.c.} \ .
\eea
Here  $O_{\rm R}^{\pm}(\mu)$ are the four-fermion operators
\bea
\label{def_Opm}
 O^{\pm} &=&\dfrac12\left\{(\bar{s}\gamma_{\mu}^L u)(\bar{u}\gamma_{\mu}^L d) 
 \pm (\bar{s}\gamma_{\mu}^L d)(\bar{u}\gamma_{\mu}^L u)\right\}
 - [u \leftrightarrow c] \ ,
\eea
renormalised at a scale $\mu$ (the subscript ${\rm R}$ 
indicates renormalised operators), $C_\pm(\mu)$ are the corresponding
Wilson coefficients, and
$\gamma_{\mu}^L = \gamma_{\mu}(1-\gamma_5)$. 
In principle, a further contribution to the weak Hamiltonian comes from
integrating out the top quark. This has been neglected
here, as it is suppressed by a factor 
$(V_{td}V^*_{ts})/(V_{ud}V^*_{us}) = \rmO(10^{-3})$.

The operators $O^\pm$ can be classified according
to isospin symmetry as follows: $O^-$ is in the  
$I=1/2$ representation, whereas 
$O^+$ is the sum of two contributions,
\be
    O^+ = O^+_{1/2} + O^+_{3/2},
 \label{isospin_decomposition}
\ee
corresponding to the representations with
$I=1/2$ and $I=3/2$ respectively.
These are given explicitly by
\bea
    O^+_{1/2} &=& \dfrac16 \Bigl[ 
   (\bar s\gamma_\mu^L u)(\bar u\gamma_\mu^L d)+ 
   (\bar s\gamma_\mu^L d)(\bar u\gamma_\mu^L u)+ 
   2(\bar s\gamma_\mu^L d)(\bar d\gamma_\mu^L d)\Bigr]\nonumber\\
   &&\mbox{}- \dfrac12\Bigl[(\bar s\gamma_\mu^L c)(\bar c\gamma_\mu^L d)
            +(\bar s\gamma_\mu^L d)(\bar c\gamma_\mu^L c)\Bigr],\\
    O^+_{3/2} &=& \dfrac13 \Bigl[ 
   (\bar s\gamma_\mu^L u)(\bar u\gamma_\mu^L d)+ 
   (\bar s\gamma_\mu^L d)(\bar u\gamma_\mu^L u)- 
   (\bar s\gamma_\mu^L d)(\bar d\gamma_\mu^L d)\Bigr].
\eea
The $\Delta I=1/2$ rule implies that
the contribution  of the $I=1/2$
component of the effective Hamiltonian to $K\to\pi\pi$ matrix elements
is much larger than that of its $I=3/2$ part. 
This contradicts naive expectations 
and results in the $1/N_c$ expansion~(cf.~\cite{lo}).

Due to the chiral structure of the weak interactions, the 
operators or the effective Hamiltonian transform as singlets
 under the ${\rm SU}(3)_{\rm R}$
part of the chiral flavour group. Their classification
according to the irreducible octet
and 27-plet representations of ${\rm SU}(3)_L$ can be found 
e.g.~in ref.~\cite{Hernandez:2002ds}. 
Here we note that we will be interested in regularisations
with Wilson quarks and non-standard mass terms, which
partially break the chiral flavour symmetries. The decomposition
of the renormalised operators into irreducible chiral representations 
is then non-trivial and hence of limited use.

\subsection{Effective chiral theory of kaons and pions}

At low energies kaons and pions are the
relevant degrees of freedom, and their interactions can be 
described by chiral perturbation theory ($\chi$PT)~\cite{Gasser:1983yg},
in terms of the (Euclidean) action
\be
    S_{\rm eff} = \int {\rm d}^4x
   \dfrac{F^2}{4}\left\{\Tr\left(\partial_\mu U\partial_\mu U^\dagger\right)
    -\Tr\left(U\chi^\dagger + \chi U^\dagger\right)\right\}.
\ee
Here, $F$ is the the pseudoscalar meson decay constant, normalised
so that $F_\pi=92.4\, \MeV$ is the experimental value determined
from pion decay. The SU(3) matrix field $U(x)$ collects 
the pion, kaon and eta fields, parameterised as 
\be
 U(x) = \exp(\sqrt{2}i\Phi(x)/F),
\ee
with
\be
\Phi = 
 \left(
 \begin{array}{ccc} 
 \pi^0/\sqrt{2}+\eta/\sqrt{6} & \pi^+                        & K^+ \\
 \pi^-                        &-\pi^0/\sqrt{2}+\eta/\sqrt{6} & K^0 \\
  K^-                         &\bar{K}^0                     &-2\eta/\sqrt{6}
 \end{array}  
 \right).
\ee
A chiral flavour transformation 
$(g_L,g_R)\in {\rm SU}(3)_L\times{\rm SU(3)}_R$ transforms the $U$-field 
according to 
\be
   U\rightarrow g_L^{} U g_R^\dagger,
\ee
and the action is  invariant under this symmetry,
provided the spurion field $\chi$ transforms in the same way as 
$U$. The symmetry is broken explicitly by setting 
\be
  \chi= 2 B {\cal M}, \qquad {\cal M}= {\rm diag}(m_u,m_d,m_s),
\ee
where $B$ is a constant. Assuming 
isospin symmetry, $m_u=m_d=\hat m$, and expanding the
$U$-field in powers of $\Phi$, one finds the lowest order
relation between meson masses and quark masses, 
\be
    m_\pi^2=2 B \hat m,\qquad m_K^2= B(m_s+\hat m),
\ee
which implies,
\be
   \chi = {\rm diag}(m_\pi^2,m_\pi^2,2m_K^2-m_\pi^2). 
\ee
The weak vertices for $\Delta S=1$ processes
which correspond to the effective 4-quark
operators in the Standard Model 
can now be identified by their transformation
behaviour under the chiral flavour symmetry. Following the
conventions of~\cite{Hernandez:2002ds} we have
\be
   {\cal H}_w(x) =2\sqrt{2}G_{\rm F}V_{ud}^{}V_{us}^\ast 
    \left\{\frac53 g_{27} {\cal O}_{27}(x) + 2 g_{8} {\cal O}_{8}(x)
           +2 g_{8}' {\cal O}_{8}'(x) \right\} + {\rm h.c.} \ .
\ee
Using the notation
\be
    L_\mu = \frac{F^2}2 \partial_\mu U U^\dagger,
\ee
the operators are explicitly given by
\bea
 {\cal O}_{27} &=& \frac35 (L_\mu)_{23}(L_\mu)_{11} 
                 + \frac25 (L_\mu)_{21}(L_\mu)_{13},\\
 {\cal O}_{8}  &=& \frac12\sum_{k=1}^3 (L_\mu)_{2k}(L_\mu)_{k3}, \\
 {\cal O}_{8}' &=& \frac14 F^4 (U\chi^\dagger+\chi U^\dagger)_{23}.
\eea
Besides the parameters $F$ and $B$, 
first order $\chi$PT for electroweak processes of $\Delta S=1$ 
contains the undetermined couplings
$g_{27}$, $g_8$ and $g_8'$. 
Moreover, it can be shown that the so-called weak mass term
proportional to $g_8'$ does not contribute to
physical kaon decay amplitudes~\cite{Bernard:wf,rc}.

By matching experimental results for $K\to\pi\pi$ decays 
to the rates obtained at lowest order chiral perturbation 
theory one may obtain a phenomenological estimate of 
$g_8$ and $g_{27}$. Taking into account experimental
information about the scattering phases between the
pions in the final state, the authors of refs.~\cite{lo,gm}
obtained the values
\be
     |g_8|   \approx 5.1,  \qquad     
     |g_{27}|\approx 0.29.
   \label{g_pheno}
\ee
The clear hierarchy between the couplings 
reflects the $\Delta I=1/2$ rule in the framework
of leading order chiral perturbation theory.

\subsection{$K\to\pi$ amplitudes}

An explanation of the $\Delta I=1/2$ rule in this framework
requires the determination of the couplings $g_8$ and $g_{27}$ 
directly from QCD. Instead of a direct determination of  
the physical $K\to\pi\pi$ transitions,
it has become customary to study the unphysical
$K\to\pi$ amplitudes~\cite{Bernard:wf}.
Using standard conventions (see e.g~\cite{Itzykson})
the matrix element for $K^+\to\pi^+$ transitions 
in the effective theory is
\be
   \left\langle \pi^+,{\bf q}|{\cal H}_w(0)|K^+, {\bf p}\right\rangle = 
    \sqrt{2}G_{\rm F}V_{ud}V_{us}^\ast F^2\left\{
     \left(\frac23 g_{27}+g_8\right) p\cdot q
     + 2m_K^2g_8'\right\},
\label{eq:piK}
\ee
where 
\be
  p\cdot q = E_K({\bf p})E_\pi({\bf q})-{\bf p}\cdot{\bf q},\qquad
  E_X({\bf p})=\sqrt{m_X^2+{\bf p}^2}.
\ee
In order to determine the combination $\frac{2}{3}g_{27}+g_8$ we note that 
this term is proportional to the product of spatial momenta
${\bf p} \cdot {\bf q}$, while the unwanted $g_8^\prime$ term is part
of the $m_K^2$ contribution. One could thus isolate the former term by
e.g.~evaluating the $K^+\to\pi^+$ matrix element in
lattice QCD for two different pion energies (i.e. two different sets
of spatial momenta). A technical disadvantage of this approach (as
opposed to the one adopted with domain wall fermions in~\cite{GW}) is
that it requires computation of lattice correlation functions with
non-zero spatial momenta, which are noisy.

In order to obtain both $g_{27}$ and $g_8$ independently
one then needs a second matrix element. If isospin symmetry
is unbroken on the lattice, one may use the decomposition
(\ref{isospin_decomposition}) and the corresponding one
in the effective theory, where the octet operator only
mediates transitions with $\Delta I=1/2$, whereas
the 27-plet operator contains both, $\Delta I=1/2$ and
$3/2$. Decomposing this operator
\be
   {\cal O}_{27}={\cal O}_{27}^{1/2}+
                 {\cal O}_{27}^{3/2},
\ee
with
\bea
   {\cal O}_{27}^{1/2} &=&\frac{1}{15}\Bigl[(L_\mu)_{21}(L_\mu)_{13}
  +(L_\mu)_{23}\left\{ 4(L_\mu)_{11}+5(L_\mu)_{22}\right\}\Bigr],\\
   {\cal O}_{27}^{3/2} &=& \frac13\Bigl[(L_\mu)_{21}(L_\mu)_{13}
  +(L_\mu)_{23}\left\{ (L_\mu)_{11}-(L_\mu)_{22}\right\}\Bigr],
\eea
it is easy to check that the $K^+\to\pi^+$ matrix elements of
these operators yield two independent combinations
of $g_8$ and $g_{27}$.
If isospin is not a lattice symmetry the isospin decomposition 
of the lattice operators poses a difficult renormalisation problem by itself.
Rather than addressing this problem, one may instead look
at the $K^0\to\pi^0$ matrix element,
\be
   \left\langle \pi^0,{\bf q}|{\cal H}_w(0)|K^0, {\bf p}\right\rangle = 
      G_{\rm F} V_{ud}V_{us}^\ast F^2\left\{
     \left(g_{27}-g_8\right) p\cdot q
     - 2m_K^2g_8'\right\},
\ee
which yields the combination $g_{27}-g_8$.
However, we anticipate that this matrix element may
be difficult to evaluate in twisted mass lattice QCD (cf.~sect.~5.6).
\par 
Finally we recall the fact that $g_{27}$ can be related, via chiral
symmetry, to $B_K$ in the chiral limit. Using 
an independent determination of $B_K$ as input, 
one may subsequently obtain $g_8$ by studying  
the $K^+\to\pi^+$ matrix element of eq.~(\ref{eq:piK}).

\subsection{Renormalisation of the four-quark operators on the lattice}

In practice the renormalisation of the four-quark operators
has been one of the major stumbling blocks 
for lattice regularisations which do not preserve
chiral symmetry, such as lattice QCD with Wilson type quarks~\cite{Gavela:ws}.
Decomposing the operators $O^{\pm}$ into parity-even and parity-odd parts, 
\bea
 \label{split_Opm}
 O^{\pm} = O^{\pm}_{\rm VV+AA} - O^{\pm}_{\rm VA+AV},
\eea
we see that parity ensures that $K\to\pi\pi$ matrix elements 
receive contributions only from
$O^{\pm}_{\rm VA+AV}$, while $K\to\pi$ matrix elements  
arise only from $O^{\pm}_{\rm VV+AA}$. 

To illustrate the importance of chiral symmetry
we briefly recall the counterterm structure
in the case of an ideal regularisation with cutoff $1/a$,
in the sense that the continuum symmetries are 
preserved~(see e.g.~\cite{Donini:1999sf}). 
In this case one finds that the operators 
renormalise as follows:
\bea
  (O_{\rm R})^{\pm}_{\rm VA+AV} = & Z^{\pm}\Big[O^{\pm}_{\rm VA+AV} 
       + c^\pm \, (m_c^2 - m_u^2) \,\, (m_s - m_d)\, \bar s \gamma_5 d 
  \Big], 
  \label{VAAV_renGW}\\
  (O_{\rm R})^{\pm}_{\rm VV+AA} = & Z^{\pm}\Big[O^{\pm}_{\rm VV+AA}  
       + c^\pm \, (m_c^2 - m_u^2) \,\, (m_s + m_d)\, \bar
       s d\Big] \ .
  \label{VVAA_renGW}
\eea
Here, the multiplicative renormalisation constants $Z^\pm$
and the additive renormalisation constants
$c^\pm$ are the same for both operators.
Counterterms which vanish by the equations of motion 
have been omitted. This is justified as long as
only on-shell correlation functions are considered. 
Note that lattice regularisations with Ginsparg-Wilson fermions
provide explicit examples for the ideal regularisation,
if one disregards the breaking of the continuous space-time symmetries. 
Therefore, with Ginsparg-Wilson fermions, 
the renormalisation pattern of the operators 
can indeed be cast in the form (\ref{VAAV_renGW},\ref{VVAA_renGW})~\rep{GW_theor}. 
Furthermore, one expects that the operators are on-shell O($a$) improved, 
i.e.~leading cutoff effects in their matrix elements are of O($a^2$).

In contrast, for Wilson-type quarks one finds the counterterm 
structure~(see e.g.~\cite{Maiani:1986db,Gavela:ws,dih_review_2})
\bea
  (O_{\rm R})^{\pm}_{\rm VA+AV} &=& Z^{\pm}\Big[O^{\pm}_{\rm VA+AV} 
               + c^{\pm}_P \,\bar s \gamma_5 d \Big],  
   \label{VAAV_renW}\\
   (O_{\rm R})^{\pm}_{\rm VV+AA} & = & Z^{\pm}\Big[
\sum_{k=1}^5  Z^{\pm}_k O^{\pm}_k + c^{\pm}_S \ \bar s d + d^{\pm}_\sigma \, 
  \bar s \sigma_{\mu\nu} F_{\mu\nu} d \Big].
\label{VVAA_renW}
\eea
Here $k\in \{\rm VV+AA, VV-AA, SS, PP, TT\}$ labels the
parity even Dirac structures of the four-quark operators 
in standard notation~\cite{Donini:1999sf}, and $F_{\mu\nu}$ denotes
some lattice version of the gluon field strength tensor.
Once more, counterterms which vanish by the equations of motion
have been neglected.

A first observation is that the parity even and odd operators
are renormalised differently. Compared to the chirally
symmetric regularisation one also finds that some of
the additive renormalisation constants are
divergent as $a\to 0$: 
\bea
  c^\pm_P &\sim& \frac{1}{a} \, (m_c - m_u) \, (m_s - m_d) \nonumber \ ,\\
  c^\pm_S &\sim& \frac{1}{a^2} \, (m_c - m_u) \nonumber \ ,\\
  d^\pm_{\sigma} &\sim& (m_c - m_u) \label{lead_Wil}.
\eea
This is due to the loss of a factor $(m_c+m_u)$ and,
for the parity even operator, of a further factor 
$(m_s+m_d)$. Instead of a finite counterterm
one now has linear and quadratic divergences for the
parity odd and even operators respectively. In the
latter case there is also a subleading finite counterterm. 
An additional complication arises in the parity even operator
due to the mixing with four other operators of dimension~6. 
Finally, we note that the renormalised operators are not O($a$) improved.
These complicated renormalisation properties of 
the parity even operator have prevented
the computation of $K\to\pi$ transitions with Wilson 
fermions~\cite{Gavela:ws}. 
On the other hand, the situation for the parity odd operator is not
much worse than the case of a chirally symmetric regularisation:
only a single additive counterterm needs to be
subtracted (besides the multiplicative renormalisation).
This motivates us to investigate whether twisted mass
QCD may alleviate the lattice renormalisation problem,
in a similar way as discussed for $F_\pi$ 
and the matrix element for $K^0$--$\bar{K}^0$ mixing in 
refs.~\cite{tmQCD_proc1,tmQCD_1,Frezzotti:2001du}.

\section{Twisted mass QCD with four quark flavours}
\label{sec:tmQCD}

Twisted mass QCD has been designed to eliminate
exceptional configurations in (partially) quenched 
lattice simulations with light Wilson quarks~\rep{tmQCD_proc1}. 
In its original formulation, it describes a
mass-degenerate isospin doublet $\psi$ of Wilson quarks for which, 
besides the standard mass term, a so-called twisted mass term
$i\mu_{\rm q}\bar{\psi}\gamma_5\tau^3\psi$ is introduced. 
The properties of tmQCD have been studied in
detail in~\rep{tmQCD_1}, where, in particular, 
its equivalence to standard two-flavour QCD has been established.
We discuss here the extension of this framework to four quark flavours.

\subsection{Classical action}
\label{subsec:action}

We start by considering tmQCD in the continuum with
the fermionic action
\be
 \label{tmQCD_action}
 S_F = \int{\rm d}^4 x \,\,\bar{\psi}(x)\left(D_\mu\gamma_\mu 
+ \smm + i\gamma_5\tmm\right)\psi(x),
\ee
where, $\psi^T = (u,d,s,c)$. The standard and twisted mass 
matrices have the form, respectively:
\bea
 \nonumber
 \smm &=& \diag\left(M_u\cos\taa,M_d\cos\taa,M_s\cos\tab,M_c\cos\tab\right) \\
 \label{standard_mass_matrix}
 &\equiv& \diag(m_u,m_d,m_s,m_c) \, , \\
 \nonumber
 \tmm &=& \diag\left(M_u\sin\taa,-M_d\sin\taa,M_s\sin\tab,-M_c\sin\tab\right)\\
 \label{twisted_mass_matrix}
 &\equiv& \diag(\mu_u,\mu_d,\mu_s,\mu_c) \, .
\eea
Hence the theory has six independent parameters, 
namely the four radial quark masses $M_i$ ($i=u,d,s,c$),
with $M_i^2 = m_i^2 + \mu_i^2$, and the two twist angles 
$\taa,\tab$. In other words, the four standard mass parameters 
$m_i$ and the four twisted mass parameters $\mu_i$ are constrained by
\be
 \tan \taa = \frac{\mu_u}{m_u} = - \frac{\mu_d}{m_d},
 \qquad
 \tan \tab = \frac{\mu_s}{m_s} = -\frac{\mu_c}{m_c}.
 \label{eq:tan}
\ee
This framework is a natural extension of two-flavour tmQCD, 
with the property that the quark mass terms remain flavour diagonal. 
At vanishing twist angles $\alpha$ and $\beta$
one recovers the standard QCD action of four quark flavours,
while for $\tab=0$ and $M_u=M_d$ the two-flavour version 
of tmQCD of ref.~\rep{tmQCD_1} is reproduced, 
with two additional untwisted quark flavours $s$ and $c$. 

The action~(\ref{tmQCD_action}) 
is form invariant under the axial field transformations
\be
 \psi    \to \psi^\prime = R(\tata,\tatb)\psi, \qquad
 \psibar \to \psibar^\prime = \psibar R(\tata,\tatb), 
 \label{axial_transf} 
\ee
with
\bea
  R(\tata,\tatb) &=& 
\exp\left\{\frac{i}{2}\gamma_5\left(\tata\tau^3_l 
 + \tatb\tau^3_h\right)\right\}, \\
 \label{tau3s}
 \tau^3_l &=& \diag(1,-1,0,0), \label{eq:tau3l} \\
 \tau^3_h &=& \diag(0,0,1,-1). \label{eq:tau3h}
\eea
The new mass term is given by
\be
 \smm^\prime + i \gamma_5 \tmm^\prime=
 R(2\tata,2\tatb) \left[\smm
 + i \gamma_5 \tmm \right],
 \label{mass_matrix_transf}
\ee
and can again be parameterised as in 
eqs.~(\ref{standard_mass_matrix},\ref{twisted_mass_matrix}).
The corresponding twist angles 
$(\taa^\prime, \tab^\prime) = (\taa-\tata,\tab-\tatb$)
are related to the mass parameters $\smm^\prime, \tmm^\prime$ 
as in~\req{eq:tan}. The radial masses $M_i$ remain invariant.  
With the choice $\tata=\taa,\tatb=\tab$ one obtains $\tmm^\prime= 0$ 
and $\smm^\prime=\diag(M_u,M_d,M_s,M_c)$, 
i.e.~four-flavour QCD with a standard mass term.
This demonstrates the equivalence of tmQCD and 
standard QCD at the level of the classical action.
It also implies that both theories share all  
the symmetries. For example, a parity transformation, 
$x\to\tilde x \equiv (x_0,-{\bf x})$, 
is realised in tmQCD by the field transformations
\be
 \psi(x)  \to \gamma_0 R\left(2\alpha,2\beta\right)\psi(\tilde x), \qquad
 \bar{\psi}(x) \to \bar{\psi}(\tilde{x})R\left(2\alpha,2\beta\right)\gamma_0,
 \label{parity_transf}
\ee
with angles $\alpha, \beta$ given by \req{eq:tan}.

\subsection{Relations between composite fields}

In general, the axial transformation (\ref{axial_transf})
induces a mapping between composite fields.
For quark bilinear operators, it can be written in a compact form:
\bea
 \label{transf_2f}
 \bar{\psi}_{i}\Gamma\psi_{j} &=& 
 \cos(\phi_{ij}^{\Gamma})\bar{\psi}^\prime_{i}\Gamma\psi^\prime_{j} +
 i \eta_\Gamma \sin(\phi_{ij}^{\Gamma})\bar{\psi}^\prime_{i}
 \Gamma\gamma_5\psi^\prime_{j}.
\eea
Here we have assumed that the spin matrices $\Gamma$ either commute 
or anticommute with $\gamma_5$. The phases are then defined by
\be
 \label{angle_combination}
 \phi_{ij}^{\Gamma} = \omega_{i} - \eta_{\Gamma}\omega_{j}, 
\qquad
 \eta_{\Gamma} = \left\{\ba{ll}+1,&{\rm~if~}\{\gamma_5,\Gamma\}=0,\\
 -1,&{\rm~if~}[\gamma_5 \ ,\Gamma]=0 \ ,\ea\right.
\ee
and
\be
\label{angle_assignment}
        \omega_u=-\omega_d=\frac{\tata}{2},
        \qquad \omega_s=-\omega_c=\frac{\tatb}{2}.
\ee
These relations establish a ``dictionary'' between composite fields
in theories with different parameterisations of the quark mass terms.
As shown in ref.~\cite{tmQCD_1} the very same relations hold between
properly renormalised fields in the quantum theories, provided 
the renormalisation scheme respects the chiral and flavour 
symmetries of the massless continuum theory. 

For illustration and later use we quote a few specific examples.
With the notation for quark bilinear fields
\be
\label{bilinears}
 S_{ij} =  \bar{\psi}_{i} \psi_{j}, \quad
 P_{ij} =  \bar{\psi}_{i} \gamma_5 \psi_{j}, \quad
 A_{\mu,ij} =  \bar{\psi}_{i} \gamma_\mu\gamma_5 \psi_{j}, \quad
 V_{\mu,ij} =  \bar{\psi}_{i} \gamma_\mu \psi_{j},
\ee
we find, for instance
\bea
  S_{us} &=& \cos\left(\tfrac{\tata + \tatb}{2}\right) S_{us}^\prime 
  - i \sin\left(\tfrac{\tata + \tatb}{2}\right) P_{us}^\prime, 
 \label{eq:Sus_transf} \\
  P_{us} &=& \cos\left(\tfrac{\tata + \tatb}{2}\right) P_{us}^\prime 
  - i \sin\left(\tfrac{\tata + \tatb}{2}\right) S_{us}^\prime,
 \label{eq:Pus_transf} \\
  A_{\mu,us} &=& \cos\left(\tfrac{\tata - \tatb}{2}\right) A_{\mu,us}^\prime 
  + i \sin\left(\tfrac{\tata - \tatb}{2}\right) V_{\mu,us}^\prime,  \\
  V_{\mu,us} &=& \cos\left(\tfrac{\tata - \tatb}{2}\right) V_{\mu,us}^\prime 
  + i \sin\left(\tfrac{\tata - \tatb}{2}\right) A_{\mu,us}^\prime.
 \label{eq:transform_ops3} 
\eea
We observe that the axial transformation amounts to a rotation
by the angle $(\tata + \tatb)/2$ in the coordinates $(P_{us},-iS_{us})$
and by the angle $(\tata - \tatb)/2$ in the coordinates $(A_{us},iV_{us})$.
With a sign flip  $\tata\to-\tata$ the same equations hold
if the up quark is replaced by a down quark.
As for operators with up and down quarks only,
we find
\be
  P_{du} = P'_{du}, \qquad  A_\mu^3  = {A'}_\mu^{3},
\ee
where we have used the isospin notation
\be
   A_\mu^3 = \frac12 \left(A_{\mu,uu}-A_{\mu,dd}\right). 
\ee

\subsection{Partial current conservation}
\label{subsec:operators}
With non-degenerate quark masses all chiral and flavour 
symmetries are broken explicitly. This is expressed
by the non-vanishing r.h.s.~of the PCAC and PCVC relations.
For the flavour non-diagonal operators with $i\ne j$,
they read,
\bea
  \partial_\mu A_{\mu,ij} &=& (m_i+m_j)P_{ij} + i(\mu_i+\mu_j)S_{ij},
 \label{PCAC}\\
  \partial_\mu V_{\mu,ij} &=& (m_i-m_j)S_{ij} + i(\mu_i-\mu_j)P_{ij}.
  \label{PCVC}
\eea
These relations take their usual form when expressed
in terms of the axially transformed fields.
Indeed, setting $\tilde\alpha=\alpha$ and $\tilde\beta=\beta$,
the PCAC and PCVC relations become
\be
  \partial_\mu A'_{\mu,ij} = (M_i+M_j)P'_{ij}, \qquad
  \partial_\mu V'_{\mu,ij} = (M_i-M_j)S'_{ij}.
\ee

\subsection{Mapping of four-quark operators}

Using eq.~(\ref{transf_2f}) we obtain the relations
for the four-quark operators of interest:
\bea
 \label{Opm_twist_1}
 O^{\pm}_{\rm VV+AA} &=& 
 \cos\left(\tfrac{\tata+\tatb}{2}\right)O^{\prime\,\pm}_{\rm VV+AA}
  +i\sin\left(\tfrac{\tata+\tatb}{2}\right)O^{\prime\,\pm}_{\rm VA+AV}, \\
 \label{Opm_twist_2} 
  O^{\pm}_{\rm VA+AV} &=& 
 \cos\left(\tfrac{\tata+\tatb}{2}\right)O^{\prime\,\pm}_{\rm VA+AV}
 +i\sin\left(\tfrac{\tata+\tatb}{2}\right)O^{\prime\,\pm}_{\rm VV+AA}.
\eea
The transformation is a rotation 
by the angle $(\tata+\tatb)/2$ in the plane spanned 
by $(O^{\pm}_{\rm VV+AA},iO^{\pm}_{\rm VA+AV})$
and can be easily inverted by reversing the sign of the twist angles. 
Setting again $\tata=\alpha$ and $\tatb=\beta$ 
it is worth noting two special cases: 
\begin{enumerate}
\item $\alpha=\beta=\pi/2$: with this choice 
the operators are directly interchanged, 
as eqs.~(\ref{Opm_twist_1},\ref{Opm_twist_2}) become
\be
 O^{\pm}_{\rm VV+AA} = i O^{\prime\,\pm}_{\rm VA+AV}, \qquad
 O^{\pm}_{\rm VA+AV} = i O^{\prime\,\pm}_{\rm VV+AA}.
\ee
In terms of the quark masses these twist angles
correspond to vanishing standard mass parameters,
\be
   m_u=m_d=m_s=m_c=0,
\ee
and positive twisted mass parameters for up and strange quarks.
\item $\alpha=-\beta=\pi/2$: in this case one
finds
\be
 O^{\pm}_{\rm VV+AA} = O^{\prime\,\pm}_{\rm VV+AA}, \qquad
 O^{\pm}_{\rm VA+AV} = O^{\prime\,\pm}_{\rm VA+AV}.
\label{Kpipi}
\ee
In terms of quark masses the only difference with respect to the previous case
is a change of sign for the twisted mass parameters of the
strange and charm quarks.
\end{enumerate}
We will refer to both these cases as ``fully twisted'', since the
physical quark masses $M_i$ are then determined by the
twisted mass parameters alone. In the fully twisted cases
the operator $O^{\pm}_{\rm VA+AV}$ can play the r\^ole
of either the parity even or the parity odd operator,
depending on the sign of $\beta$.

\subsection{Relations between correlation functions}

Consider, in the formal continuum theory, 
Euclidean correlation functions of the form
\be
\left\langle O[\psi,\psibar]\right\rangle_{(\alpha,\beta)} = 
{\cal Z}^{-1}\int_{\rm fields} O[\psi,\psibar]\rme^{-S},
\ee
where $O[\psi,\psibar]$ denotes some
multilocal gauge invariant field. The fermionic part of 
the action is given in~(\ref{tmQCD_action}).
Performing the field transformation~\req{axial_transf} and changing
integration variables in the functional integral one derives
the identity
\be
   \left\langle O\left[
   R(\tata,\tatb)\psi,\psibar R(\tata,\tatb)\right]
   \right\rangle_{(\alpha,\beta)}
 = \left\langle O\left[\psi,\psibar \right]
   \right\rangle_{(\alpha-\tata,\beta-\tatb)}
\ee
where the integration variables are $\psi$ and $\psibar$ on
both sides of the equation, and the Jacobian is unity, due to
\be
    \det\left(R(\tata,\tatb)\right)=1.
\ee
Setting $\tata=\alpha$ and $\tatb=\beta$ we thus
obtain identities between standard QCD and twisted mass QCD correlation
functions, which we can use in two alternative versions:
\bea
   \left\langle O\left[\psi,\psibar\right]
   \right\rangle_{(\alpha,\beta)}
 &=& \left\langle O\left[R(-\alpha,-\beta)\psi,\psibar R(-\alpha,-\beta)\right]
   \right\rangle_{(0,0)},
  \label{identity_1}\\
   \left\langle O\left[\psi,\psibar\right]
   \right\rangle_{(0,0)}
 &=& \left\langle O\left[R(\alpha,\beta)\psi,\psibar R(\alpha,\beta)\right]
   \right\rangle_{(\alpha,\beta)}.
 \label{identity_2}
\eea
Hence, a given correlation function in tmQCD with twist
angles $(\alpha,\beta)$ is interpreted as the linear
combination on the r.h.s.~of \req{identity_1}. If instead we
are given a standard QCD correlation function, then
\req{identity_2} tells us how it is represented in tmQCD at
twist angles $(\alpha,\beta)$.
The dictionary just established for composite fields can be
used for the integrands in \req{identity_1}, provided $\psi'$ and
$\psibar'$ are identified with the integration variables on the r.h.s..
As shown in ref.~\cite{tmQCD_1}, the relations derived 
for the formal continuum theory are realised 
between the renormalised quantum field theories, provided 
the renormalisation procedure is set up with some care.

\section{Lattice regularised tmQCD}
\label{subsec:lattice}

To regularise tmQCD on the lattice, 
we assume that the gauge fields are represented
by some standard lattice action (e.g. Wilson's plaquette action),
whereas quarks are described by the lattice action
\be
   S_{\rm f}= a^4\sum_x \psibar(x)\left(D_W + \smm 
   + i\gamma_5\tmm\right)\psi(x),
 \label{lataction}
\ee
with $D_W$ being the standard, possibly O($a$) improved 
Wilson-Dirac operator (for unexplained notation see \cite{Luscher:1996sc}).
In order to ensure the equivalence 
between tmQCD and standard QCD, the bare mass parameters, 
$m_{i}$ and $\mu_i$ ($i=u,d,s,c$),
will be subject to constraints. However, these are now only
implicitly defined by eqs.~(\ref{eq:tan}), as they will
be imposed on the properly renormalised mass parameters.
Before discussing renormalisation of lattice tmQCD
we consider the possible choices for the bare mass parameters 
leading to feasible numerical simulations.

\subsection{Quark determinant and practical choices of parameters}
\label{practical}

Whether a given lattice action is suitable for
numerical simulations obviously depends on the available
algorithms. The answer may therefore change with time.
At present, the starting point for most algorithms 
is the functional integral after integration over the quark fields. 
The resulting determinant is then part of the 
effective gauge field measure and is generally required to 
be real and positive. In the case of lattice tmQCD with four quark 
flavours as introduced here, each flavour
contributes a factor to the determinant. 
Assuming the light doublet to be mass degenerate,
\be
    m_u=m_d=m_l,\qquad  \mu_u=-\mu_d=\mu_l,
\ee
the determinants of the two light flavours can be combined
to yield
\be
 \det\left(D_W + m_l + i\mu_l\gamma_5\tau^3\right)=
 \det\left[ \left(D_W + m_l \right)^\dagger \left(D_W + m_l \right) 
  + \mu_l^2 \right],
\ee
where the determinant on the r.h.s.~is taken in a single 
flavour space~\rep{tmQCD_1}.
Hence the determinant of the light doublet is positive at non-zero 
$\mu_l$, irrespective of the background gauge field. 
Integrating over strange and charm quarks one obtains
\bea
 \nonumber
 &&
 \det\Bigl[\left(D_W + m_s \right)^\dagger
 \left(D_W + m_c \right) - \mu_s \mu_c \\
 &&
 \mbox{}+ i \, \mu_c \gamma_5 \left\{D_W + m_s \right\}  
 + i \, \mu_s \gamma_5 \left\{D_W + m_c \right\}  \Bigr],
\eea
which, in general, is not a real quantity. Discarding
the unphysical situation of mass-degenerate strange and charm quarks,
the only way to ensure the reality of the determinant is to
employ untwisted strange and charm quarks, $\mu_s=\mu_c=0$,
as the fermion determinants for individual Wilson quark flavours are real.
However, positivity of the determinant is then 
not guaranteed, unless the quark masses are large enough.
Detailed statements depend on the lattice gauge action and
the chosen simulation parameters, but for practical
purposes the strange quark mass 
might be heavy enough~(see, e.g.~\cite{Aoki:2003re}).

In view of this situation we distinguish the 
following practical options for lattice tmQCD simulations: 
\begin{itemize}
\item Quenched simulations or partially quenched simulations 
with $\Nf=2$ dynamical light and mass degenerate quarks.
Strange and charm quarks remain quenched and
do not contribute to the determinant, which is therefore guaranteed
to be real and positive (for $\mu_l\ne0$).
Hence, one has almost complete freedom in the choice of the 
mass parameters and therefore of $\alpha$ and $\beta$.
In the quenched approximation one would just require
a non-zero twist angle $\alpha$ in order to ensure the absence 
of unphysical zero modes.

\item $\Nf=3,4$ dynamical quark flavours with two mass degenerate
light quarks, and untwisted strange and charm quarks. This implies
the choice for the second twist angle $\beta=0$, 
while $\alpha$ remains unrestricted.
\end{itemize}
Note that these considerations pertain to our specific 
choice of flavour diagonal mass terms. If one allows
for flavour non-diagonal mass terms one may
indeed be able to obtain positive and real fermion determinants
for any combination of non-zero quark masses, including 
non-degenerate light quarks~\cite{Frezzotti:2003xj}. However,
this renders the relation between standard QCD and tmQCD correlation functions
more complicated, and the flavour structure has to 
be dealt with explicitly in the numerical computation of quark propagators.

\subsection{Renormalisation of lattice tmQCD}

Renormalisability is based on general properties 
of Quantum Field Theory, such as locality and unitarity, 
which may not hold in the (partially) quenched approximation of QCD.
In this paper we do not try to solve this general problem.
Instead we take the naive point of view that the (partially)
quenched theory is well-defined and can be 
obtained by suppressing the fermion
determinant with respect to some or all flavours. This is assumed
to incur no drastic change to the theory's renormalisability.
While we discuss the counterterm structure for the full theory 
with $\Nf=4$ dynamical quarks, we occasionally indicate modifications
which should be expected in the (partially) quenched case.

\subsubsection{Symmetries of the lattice action}
\label{subsec:symm}

When the mass terms are taken to vanish, the tmQCD lattice action
reduces to the usual (possibly O($a$) improved) 
Wilson action for $\Nf=4$ quarks.
Hence, the {\em massless} tmQCD lattice 
action ($\smm = \tmm = {\bf 0}$) is invariant under U(4) 
flavour rotations,
\be
   \psi \to V\psi,\qquad \psibar \to \psibar V^\dagger,
   \label{flavour-trafo}
\ee
as well as under discrete transformations, such 
as parity ($x\to \tilde x =(x_0,-{\bf x})$),
\be
   \psi(x) \to \gamma_0 \psi(\tilde x),\qquad \psibar 
   \to \psibar(\tilde x)\gamma_0, 
\label{eq:lpar}
\ee    
and charge conjugation
\be
   \psi(x) \to \cC^{-1} \psibar(x)^T,\qquad \psibar 
   \to - \psi(x)^T \cC
\label{eq:cc}
\ee
(the charge conjugation matrix satisfies
$\gamma_\mu^\ast = - \cC \gamma_\mu \cC^{-1}$).\footnote{We have not
indicated the parity and charge conjugation transformations of the
gauge field, as they are never used in this work.}

Once the mass and twisted mass terms are added, some of these
symmetries are lost. The introduction of the $\smm$ term
implies loss of the $U(4)$ flavour symmetry, unless $\smm \propto
1$. Upon introducing the $\tmm$ term this symmetry
is also lost. A reduced symmetry survives however,
if $\tmm \propto {\rm diag} (1, -1, 1, -1)$. This consists of
isospin rotations of the $SU(2)$ subgroups, along the generators 
$\tau_l^3$ and $\tau_h^3$ defined in eqs.~(\ref{eq:tau3l}) 
and (\ref{eq:tau3h}).
Charge conjugation is unaffected by the mass terms, while parity is
clearly broken by the twisted mass term.\footnote{The reader should
not be misled by the fact that the redefined parity
transformation (\ref{parity_transf}) is a symmetry of the classical
action. Although the $\tmm$ term is invariant under this symmetry, the Wilson
term is not.}

Upon adding the two mass terms, one may determine the counterterm
structure by treating $\smm$ and $\tmm$ as spurion fields
which transform under these symmetries so that 
the massive theory remains formally invariant. In particular,
under a U(4) flavour transformation~(\ref{flavour-trafo}) 
one assumes that
\be
   \smm \to V\smm V^\dagger,\qquad  \tmm \to V\tmm V^\dagger, 
   \label{spurion-flavour-trafo}
\ee
while invariance under the parity transformation (\ref{eq:lpar})
is obtained by assuming
\be
  \smm\to \smm\,, \qquad \tmm\to -\tmm\,.
  \label{eq:lmpar}
\ee
However, we stress again that in tmQCD with Wilson quarks and
nonvanishing twisted masses
{\em physical} parity, as defined in~\req{parity_transf}, 
is broken at O($a$), and is recovered only in the continuum limit.
Analogous considerations hold in the case of the U(4) vector flavour
symmetry.

It is sometimes useful to restrict attention 
to a subset of the general flavour transformations~(\ref{flavour-trafo}), 
such as flavour exchanges
\be
    \psi_i \leftrightarrow \psi_j,  \qquad 
 \psibar_i \leftrightarrow \psibar_j,
\ee
for which the spurion transformations~(\ref{spurion-flavour-trafo})
reduce to an exchange of the corresponding 
mass parameters, 
\be
    m_i \leftrightarrow m_j,\qquad \mu_i\leftrightarrow\mu_j.   
\ee
\subsubsection{Renormalised parameters}

The physical interpretation of tmQCD hinges on the knowledge of 
the twist angles in the renormalised theory. 
As these are determined by ratios of the renormalised standard and
twisted mass parameters we first discuss 
renormalisation of the bare parameters in the action.

The counterterms to the action of dimension $\leq 4$ are,
\be
    \tr\left\{F_{\mu\nu}F_{\mu\nu}\right\},\quad
     \psibar\psi, \quad 
     \tr(\smm)\psibar\psi,\quad 
     \psibar\smm\psi,\quad 
    i\tr(\tmm)\psibar\gamma_5\psi,\quad
     i\psibar\tmm\gamma_5\psi,
\label{d4counter}
\ee
where $F_{\mu\nu}$ denotes the gluon field strength tensor.
The first counterterm implies a multiplicative renormalisation 
of the bare gauge coupling. As for the quark mass renormalisation 
the situation is complicated by the trace
terms, which would be absent in a chirally symmetric
regularisation. For standard Wilson fermions the situation has been discussed 
in refs.~\cite{Bhattacharya:1999yg,Bhattacharya:2003nd}, 
which can be summarised and extended as follows: 
first one decomposes the mass matrices in non-singlet
and singlet pieces,
\be
    \smm =    \sum_{a=1}^{15}m^a  \lambda^a +m^0\unit_4, \qquad
    \tmm =    \sum_{a=1}^{15}\mu^a\lambda^a +\mu^0\unit_4,  
\ee
where $\lambda^a$, $a=1,\ldots 15$ are the generators of SU(4).
This decomposition applies to general Hermitean $4\times 4$ matrices, but
with our choice of flavour diagonal mass matrices 
only 3 of the non-singlet coefficients are non-zero.
These correspond to the 3 diagonal and traceless generators of SU(4)
and are therefore linear combinations of (twisted) mass differences.
The singlet coefficients are given by
\be
   m^0 = \frac{1}{4} \tr(\smm),\qquad \mu^0 = \frac{1}{4} \tr(\tmm) \,\, ,
\ee
The non-singlet terms are renormalised by the inverse 
renormalisation constant of the non-singlet scalar 
and pseudoscalar densities, $S_{ij}$ and $P_{ij}$ ($i\ne j$) 
respectively, 
\be
     m^a_{\rm R} = Z_{\rm S}^{-1} m^a,\qquad
    \mu^a_{\rm R}= Z_{\rm P}^{-1}\mu^a.
\ee
The renormalised singlet masses $m^0$ and $\mu^0$ are 
then given by
\be
     m^0_{\rm R}= Z_{{\rm S}^0}^{-1}(m^0-m_{\rm cr}),\qquad
   \mu^0_{\rm R}= Z_{{\rm P}^0}^{-1}\mu^0,      
\ee
where $Z_{{\rm S}^0,{\rm P}^0}$ are the multiplicative
renormalisation constants of the singlet densities 
$S^0=\sum_{i=1}^4 S_{ii}$ and 
$P^0=\sum_{i=1}^4 P_{ii}$ respectively,
and $m_{\rm cr}$ denotes the ``critical mass''.

From this renormalisation pattern we obtain for
quark mass parameters of individual flavours
\bea
\label{eq:smren}
    m_{{\rm R},i} &=& Z_{{\rm S}}^{-1}\left[m_{i}-m_{\rm cr} 
    +(r_m-1)\left(m^0-m_{\rm cr}\right)\right],\\
    \mu_{{\rm R},i} &=& Z_{{\rm P}}^{-1}\left[\mu_i 
    +(r_\mu-1)\mu^0 \right],
\label{eq:tmren}
\eea
with
\be
   r_m = Z^{}_{{\rm S}}Z_{{\rm S}^0}^{-1},\qquad   
   r_\mu = Z^{}_{{\rm P}}Z_{{\rm P}^0}^{-1}.
\ee
Here, the ratios $r_m,r_\mu$ are finite functions of the gauge
coupling, which can in principle be determined by axial Ward identities
in the chiral limit.
A few comments are in order, in relation to the practical considerations
made in subsect.~\ref{practical}:
\begin{itemize}
\item
The above counterterm structure
applies to the complete theory with $\Nf=4$ dynamical quarks, for which our
preferred choice of twist angles is $\alpha=\pi/2$ and 
$\beta=0$ (partial twist).
The latter condition translates to renormalised twisted parameters
\be
  \mu_{{\rm R},s} = \mu_{{\rm R},c}=0 \,\, ,
\ee
which is achieved by setting
\be
    \mu_s=\mu_c=0 \,\, .
\ee
Due to $\mu_u=-\mu_d=\mu_l$, this implies $\mu^0=0$
and thus no flavour mixing in the twisted mass renormalisation 
pattern expressed by eq.~(\ref{eq:tmren}). On the other hand, 
the condition concerning the twist angle $\alpha$ translates into
\be
  m_{{\rm R},l} =0 \,\, ,
\ee
which, due to eq.~(\ref{eq:smren}) implies that
the bare mass of the light quarks must be tuned as follows,
\be
   m_l = m_{\rm cr} 
   +\frac{(1-r_m)}{2(1+r_m)}\left(m_s+m_c-2m_{\rm cr}\right),
\label{tuning}
\ee
i.e.~there is an offset of O($1$) to the linear divergence
cancelled by $m_{\rm cr}$. 
\item
The ratios $r_m$ and $r_\mu$ depend implicitly on the
number of dynamical quark flavours. In particular,
their quenched values are $r_m=r_\mu=1$, 
i.e.~the flavour
mixing terms in eqs.~(\ref{eq:smren}) and
(\ref{eq:tmren}) are absent in this case~\cite{Bhattacharya:1999yg}. 
Hence, the fully twisted cases $\alpha = \pm \beta = \pi/2$ 
are obtained by just setting all standard bare masses equal to 
the critical mass $m_{\rm cr}$.
\item
In the partially quenched case it is obvious 
that valence quark masses do not
participate in the renormalisation of other quark flavours.
More specifically, for the case of interest with $\Nf=2$ light
dynamical flavours, the mass renormalisation pattern reduces to
\bea
   m_{{\rm R},l} &=& Z_{{\rm S}}^{-1} r_m
    \left(m_l-m_{\rm cr}\right) \,\, , \\
    m_{{\rm R},s,c} &=& Z_{{\rm S}}^{-1}\left[m_{s,c}-m_{\rm cr} 
    +(r_m-1)\left(m_l-m_{\rm cr}\right)\right] \,\, , \\
    \mu_{{\rm R},i} &=& Z_{{\rm P}}^{-1} \mu_i \,\, .
\label{eq:tmrenPQ}
\eea
This implies that the fully twisted case,
$\alpha = \pm \beta = \pi/2$, requires a quenched-like
tuning $m_l=m_s=m_c=m_{\rm cr}$, while for the partially twisted case,
$\alpha = \pi/2, \beta = 0$, the requirement is $m_l = m_{\rm cr}$ and 
$\mu_s = \mu_c =0$.
\end{itemize}

\subsubsection{Determination of the twist angles $\alpha$ and $\beta$}
\label{determination_alphabeta}

In practice, the twist angles are defined by ratios of the renormalised
quark masses appearing in the renormalised 
PCAC and PCVC relations~\cite{tmQCD_1}.
In general, one therefore needs to discuss the renormalisation
of the currents and densities which appear in
the PCAC and PCVC relations~(\ref{PCAC},\ref{PCVC}).
If we restrict attention to flavour off-diagonal 
quark bilinear operators ($i\ne j$), the symmetries
imply a multiplicative renormalisation in all cases, 
\be
   (A_{\rm R})_{\mu,ij}= Z_{\rm A} A_{\mu,ij},\quad   
   (V_{\rm R})_{\mu,ij}= Z_{\rm V} V_{\mu,ij},\quad 
   (P_{\rm R})_{ij}= Z_{\rm P} P_{ij},\quad     
   (S_{\rm R})_{ij}= Z_{\rm S} S_{ij}\,,   
\ee
where the bare fields are defined 
as in the classical continuum theory, \req{bilinears}.
It is well-known that the scale independent renormalisation 
constants $Z_{\rm A}$ and $Z_{\rm V}$ as well as the 
ratio $Z_{\rm P}/Z_{\rm S}$ can be 
determined\footnote{The determination of $Z_{\rm V}$ could be
avoided by using the point split vector current 
which follows from the exact flavour symmetry 
of mass degenerate standard Wilson quarks.} 
by imposing axial Ward identities as normalisation 
conditions~\cite{Bochicchio:1985xa}.
To obtain the scale-dependent renormalisation constants $Z_{\rm P}$ or
$Z_{\rm S}$, it is then sufficient to impose a further
quark mass independent renormalisation condition on one of the densities.
The choice is largely arbitrary and does not
affect the determination of the twist angles as
the renormalisation constant cancels in quark mass ratios.
Given the renormalised quark bilinear fields, and
a choice of bare parameters, the 6 renormalised mass parameters
(2 for the mass degenerate light quarks and 4 for strange and charm quarks)
can be obtained  by solving a system of 6 independent 
equations, which follow from the renormalised PCAC and PCVC
relations for different flavour combinations. 

Fortunately the situation is much simpler in
the cases of practical interest, namely the choices
$\alpha=\pm\beta=\pi/2$ and $(\alpha,\beta)=(\pi/2,0)$,
which we now discuss in more detail: 
\begin{itemize}
\item  To obtain $\alpha=\pm\beta=\pi/2$
   all renormalised standard quark
   masses should vanish, $m_{{\rm R},l} =m_{{\rm R},s}=m_{{\rm R},c}=0$,
  which implies that all bare standard quark masses are
  equal. The corresponding  critical bare mass parameter $m_{\rm cr}$ 
  can be obtained by requiring 
  \be
     \partial_\mu A_{\mu,ud} = 0,
   \label{conserv_axial}
  \ee
for some correlation function of the
axial current. Note that the renormalisation constant $Z_{\rm A}$ is not
needed here, and the angles $\alpha=\pi/2$ and $\beta=\pm\pi/2$ are 
now fixed up to a sign, which is determined by
the sign of the bare twisted mass parameters.
Moreover, the fully twisted cases are special because
the values of the renormalised twisted masses are not
needed for the definition of the twist angles.
\item
The choice $\beta=0$ is easily realised by just
setting $\mu_s=\mu_c=0$, as the counterterm structure then
implies $\mu_{\rmR,s}=\mu_{\rmR,c}=0$.
Setting the twist angle $\alpha=\pi/2$ is then still achieved
by requiring~(\ref{conserv_axial}). However, the
counterterm structure (\ref{tuning}) implies that this determination 
of $m_l$ must be repeated every time the parameters $m_s$ and $m_c$ 
are changed.
\end{itemize}
Finally we remark that the definition of renormalised quark masses 
through the PCAC and PCVC relations is convenient, as it
can be applied both in the (partially) quenched and in the
full theory. In this way one may bypass the 
discussion of the quark mass counterterms 
in the partially quenched case.

\subsection{Renormalisation pattern of $O_{\rm VA+AV}^\pm$}

In view of the $\Delta I=1/2$ amplitudes, we are interested
in the counterterm structure of the operator $O^\pm_{\rm VA+AV}$
in the presence of twisted mass terms.
We apply again a spurion analysis. In particular, 
the flavour structure of the operator suggests
to consider the two flavour exchange symmetries
\bea
 \label{switch_sd}
 & d \leftrightarrow s,\quad m_d \leftrightarrow m_s,\quad \mu_d 
 \leftrightarrow \mu_s , \\
 \label{switch_uc}
 & u \leftrightarrow c,\quad m_u \leftrightarrow m_c,\quad \mu_u 
 \leftrightarrow \mu_c ,
\eea
combined with charge conjugation and spurionic parity discussed
in sect.~\ref{subsec:symm}.
Power counting and the behaviour under these symmetry transformations
imply the following renormalisation pattern:
\be
 (O_{\rm R})^{\pm}_{\rm VA+AV} =  Z^{\pm}\left\{O^{\pm}_{\rm VA+AV} + 
 c^{\pm}_P\, P_{sd} +  c^{\pm}_S\, S_{sd}\right\}.
 \label{Opm_ren}
\ee
Note that, compared to standard Wilson quarks, 
the only change consists in the addition of the scalar density. 
The symmetries also determine the behaviour of the
coefficients in the small $a$ expansion. For the leading terms 
we find:
\bea
  c^\pm_P &=& \frac{1}{a} \left\{ (m_c-m_u)(m_s-m_d) c^{\pm}_{P,a}  +
  (\mu_c-\mu_u)(\mu_s-\mu_d) c^{\pm}_{P,b} \right\} +\rmO(1)
  \label{c_Plead},\\[1ex]
  c^\pm_S &=& \frac{1}{a} \left\{ (m_c-m_u)(\mu_s-\mu_d) c^{\pm}_{S,a} +
  (\mu_c-\mu_u)(m_s-m_d) c^{\pm}_{S,b} \right\}\label{c_Slead} +\rmO(1),\\[0ex]
  \nonumber
\eea
where the coefficients on the r.h.s.~are functions 
of the bare coupling only. We observe that both coefficients
$c^\pm_{P,S}$ are linearly divergent in general. 
Simplifications occur
in special cases: if down and strange quarks are taken to 
be mass degenerate, $m_s=m_d$, $\mu_s=\mu_d$, then both
coefficients vanish identically due to the exact symmetry
under a charge conjugation combined with the flavour exchange 
symmetry, $s\leftrightarrow d$. 
Another important simplification occurs in the fully twisted
cases discussed above, since
\be
  m_u=m_d=m_s=m_c \quad \Rightarrow\quad  c^\pm_S = {\rm O}(1),
\ee
so that one is left with the linear divergence in $c^\pm_P$,
which takes the form
\be
  c^\pm_P = \frac{1}{a} (\mu_c-\mu_u)(\mu_s-\mu_d) c^{\pm}_{P,b} +\rmO(a),
\ee
i.e.~subleading terms of $\rmO(1)$ are found to be absent.

\subsection{O($a$) improved action for fully twisted quarks}

While systematic O($a$) improvement of matrix elements of
four-quark operators is beyond the scope of this paper, 
the discussion of the fully twisted case in section~5 shows
that some simplifications occur if the action is O($a$) improved.
To investigate the structure of the O($a$) counterterms
in the special case of the fully twisted tmQCD action, we
make use of results of ref.~\cite{tmQCDimprove}, which were
obtained for two-flavour tmQCD with degenerate quarks.
These results are directly relevant here, since, as discussed
in subsection~4.1, a full twist of
all quarks is implemented either in the quenched approximation
or in the partially quenched case
with dynamical (and mass degenerate) up and down quarks.

In the following it is assumed that the massless 
theory has been O($a$) improved by the addition of the
Sheikholeslami-Wohlert term to the action
and of the counterterm proportional to $c_{\rm A}$ to the 
axial current~\cite{Luscher:1996sc}.
Let us now suppose that the fully twisted case has been realised
by tuning the standard bare masses to their critical value
as explained in subsection~\ref{determination_alphabeta}.
The O($a$) effects in the renormalised light mass parameters
can then be inferred from  the result of 
ref.~\cite{tmQCDimprove}, as the quenched strange and charm quarks
may not generate any additional counterterms:
\bea
   m_{\rmR,l} &=&  Z^{-1}_{S^0}\tilde{b}_{m_l}a\mu_l^2 +\rmO(a^2),\\
  \mu_{\rmR,l}&=& Z_{\rmP}^{-1}\mu_l +\rmO(a^2).
\eea
To find the counterterm structure for the strange
and charm mass parameters, we use the
fact that both quarks remain in the quenched approximation.
Hence, one may imagine that either quark 
is a member of a mass degenerate quark doublet, with partners
that do not enter any of the correlation functions, and are
therefore completely decoupled from the theory.
The renormalisation for each doublet then proceeds independently
with the result ($i=s,c$),
\bea
   m_{\rmR,i} &=& Z^{-1}_{S^0}\left\{\tilde{b}_{m_i}a\mu_i^2 
   +\tilde{b}'_{m_i}a\mu_l^2\right\}.\\
   \mu_{\rmR,i}&=& Z_{\rmP}^{-1}\mu_i,
\eea
where the new coefficients $\tilde{b}'_{m_i}$ are non-zero only if
the light quarks are dynamical.
At this point we recall that in the two flavour theory
the set of O($a$) mass counterterms 
as determined by the lattice symmetries 
is over-complete~\cite{tmQCDimprove}.
More precisely, an axial rotation re-parameterises the
mass counterterm basis, implying that there is
a one-parameter family of equivalent O($a$) improved
theories. This freedom may then be used to set one of the
improvement coefficients to a fixed value.
Hence, in the fully quenched case we may now choose to set
\be
  \tilde{b}_{m_l}= \tilde{b}_{m_s}=\tilde{b}_{m_c}=0,
\ee
and, as a result, all the renormalised quark mass parameters 
are O($a$) improved. In the partially quenched
case we may set $\tilde{b}_{m_l}= 0$, and for $i=s,c$,
\be
    \tilde{b}_{m_i}= -\tilde{b}'_{m_i}\mu_l^2/\mu_i^2.
\ee
We conclude that in the fully twisted cases, there is a choice for the
O($a$) mass counterterms such that the action is 
O($a$) improved, provided that 
the action and the axial current have been O($a$) improved in 
the chiral limit.

\section{$K\to\pi$ transitions from twisted mass lattice QCD}
\label{sec:correl}

We now show how to extract $K^+\to\pi^+$ matrix elements
from tmQCD correlation functions. We will mostly
be concerned with the determination of the additive counterterms to the
renormalised operator, or with corresponding subtractions 
of its correlation functions. The multiplicative operator renormalisation 
can be performed using standard 
techniques~\cite{Martinelli:1994ty,Jansen:1995ck,Donini:1999sf,PenaBK} and 
does not pose any further conceptual problems. 

In what follows we will refer to QCD with a standard
parameterisation of the quark mass term as ``standard QCD'', as
opposed to QCD with a twisted mass term. Their
respective renormalised correlation functions are thus
related by a chiral rotation of the fields, according to the
discussion in~\res{sec:tmQCD}.

\subsection{From correlation functions to matrix elements}

In standard QCD, the $K^+\to\pi^+$ matrix elements can be 
obtained from the Euclidean 3-point function 
\be 
  G^\pm_{K\pi}(x,y,z)= \left\langle \varphi_\pi(x) 
 (O_{\rm R})_{\rm VV+AA}^\pm(y) 
  \varphi_K(z)\right\rangle_{(0,0)}, 
\label{eq:GKpi}
\ee 
where $\varphi_\pi$ and $\varphi_K$ are interpolating fields with 
the quantum numbers of the charged pion and kaon,
respectively. Simple examples are the local fields, 
\be 
   \varphi_\pi(x) = P_{du}(x),\qquad \varphi_K(z)=P_{us}(z), 
   \label{local_sources}
\ee
but more complicated choices are possible and indeed often
necessary to obtain a signal in numerical simulations.
Based on the transfer matrix formalism, the spatial Fourier transform 
\be
   \widetilde{G}_{K\pi}^\pm(x_0,z_0;{\bf p,q})= 
   a^6\sum_{\bf x,z} {\rm e}^{i(\bfp\bfx -\bfq\bfz)}G^\pm_{K\pi}(x,0,z)
\ee
can then be shown to behave, for $x_0\rightarrow \infty$
and $z_0\rightarrow -\infty$, as
\bea
   \widetilde{G}^\pm_{K\pi}(x_0,z_0;{\bf p,q})
&\sim&   Z_\pi Z_K \exp\left(-E_\pi({\bf p})x_0+E_K({\bf q})z_0\right)
\nonumber\\[1ex]
&&  \times 
  \left\langle\pi^+,{\bf p}\vert (O_{\rm R})_{\rm VV+AA}^\pm(0)\vert K^+,{\bf q} 
   \right\rangle.
\eea
Here, neglected terms are exponentially suppressed contributions
of higher intermediate states in either the pion or the kaon channel.
The constants $Z_\pi$ and $Z_K$, as well as 
the pion and kaon energies $E_\pi(\bfp)$ and $E_K(\bfq)$ 
can be determined from the 2-point functions
of the interpolating kaon and pion fields, so that one can isolate
the desired matrix element of the renormalised operator. 
In fact the matrix element itself is used to completely
specify the renormalised operator, by
imposing the condition
\be
    \left\langle\pi^+,{\bf 0}\vert (O_{\rm R})_{\rm VV+AA}^\pm(0)\vert K^+,{\bf 0} 
   \right\rangle =0.
  \label{matrix_renorm}
\ee
Note that, even with a chirally
invariant regularisation, such a condition 
may be used in order to determine the O($1$) subtraction coefficients 
$c^\pm$ in \req{VVAA_renGW}.\footnote{Instead of the above condition,
in simulations with Ginsparg-Wilson quarks it is customary to determine the
subtraction coefficient by imposing that the operator matrix element of the
unphysical $K \to 0$ transition (with the $K$-meson at rest) vanishes.}
This arbitrariness is closely related to the fact that 
the $K\to\pi$ matrix elements are not directly linked to a
physical observable. Indeed, it is easy to see
that the subtracted counterterm does not contribute to
physical $K\to\pi\pi$ amplitudes. 
This situation is reflected in chiral perturbation theory,
where the condition (\ref{matrix_renorm})
determines the unphysical coupling $g_8'$.
It is actually convenient to impose
the renormalisation condition (\ref{matrix_renorm}),
as it facilitates the matching to chiral perturbation theory.

\subsection{Renormalised correlation function in tmQCD}
\label{subsec:corrls}

In the twisted mass QCD framework, we 
expect that the properly renormalised correlation functions are
mapped onto their standard QCD counterpart in the same way as in the classical
continuum theory~\cite{tmQCD_1}. According to Sect.~3 we can thus write
\bea
 G_{K\pi}(x,y,z)&=& \Bigl\langle \tilde{\varphi}_\pi(x) 
\Bigl[\cos\left(\tfrac{\alpha+\beta}{2}\right)  (O_\rmR)_{\rm VV+AA}^\pm(y)
\nonumber\\
&&\hphantom{0123456}
  -i\sin\left(\tfrac{\alpha+\beta}{2}\right)(O_\rmR)_{\rm VA+AV}^\pm(y) 
 \Bigr] \tilde{\varphi}_K(z)\Bigr\rangle_{(\alpha,\beta)}, 
 \label{G_match}
\eea
where $\tilde{\varphi}_\pi$ and $\tilde{\varphi}_K$ denote
the transformed source fields, which can be worked
out once a specific choice of pseudoscalar sources
$\varphi_\pi$ and $\varphi_K$ has been made.
For instance, to reproduce the local source
fields of eq.~(\ref{local_sources}) up to terms of O($a$), one defines
\be
    \tilde{\varphi}_K = \cos\left(\tfrac{\alpha+\beta}{2}\right)
    (P_\rmR)_{us}+i\sin\left(\tfrac{\alpha+\beta}{2}\right)(S_\rmR)_{us},
\qquad
    \tilde{\varphi}_\pi = (P_\rmR)_{du}.
\label{tw_local_sources}
\ee
Note that for generic twist angles both parity components of $O^\pm$
appear in the r.h.s.~of (\ref{G_match}), while we would like to be left
only with the operator $O_{\rm VA+AV}^\pm$, due to its 
nicer renormalisation properties.
We therefore turn the tables and determine the physical 
interpretation of the tmQCD correlation 
functions with this operator alone:
\bea
  \Bigl\langle \widetilde{\varphi}_\pi(x) (O_\rmR)_{\rm VA+AV}^\pm(y)
   \widetilde{\varphi}_K(z)\Bigr\rangle_{(\alpha,\beta)}
&=& \Bigl\langle \varphi_\pi(x) 
\Bigl[\cos\left(\tfrac{\alpha+\beta}{2}\right)  (O_\rmR)_{\rm VA+AV}^\pm(y) 
\nonumber\\
&&\mbox{}+i\sin\left(\tfrac{\alpha+\beta}{2}\right)(O_\rmR)_{\rm VV+AA}^\pm(y) 
 \Bigr] \varphi_K(z)\Bigr\rangle_{(0,0)} 
\nonumber \\
&=& i\sin\left(\tfrac{\alpha+\beta}{2}\right) G^\pm_{K\pi}(x,y,z).
 \label{G_matchII}
\eea
Here, the last step is trivial in the fully twisted case, 
$\alpha=\beta=\pi/2$, for which the cosine 
in the first term on the r.h.s.~vanishes.
On the other hand, for generic values of the 
twist angles the cosine multiplies 
a standard QCD correlation function which vanishes
by parity. Hence, by computing the l.h.s.~of eq.~(\ref{G_matchII})
in twisted mass QCD one obtains the desired correlation function
up to a known factor, and the $K\to\pi$ matrix element
is then obtained in the standard manner.
It is important at this point, however, to stress that
eq.~(\ref{G_matchII}) holds exactly only in the continuum
limit, and assumes that physical parity, defined
through~(\ref{parity_transf}), is
properly restored in renormalised tmQCD. 
In principle this can be achieved in two 
ways: given the twist angles, it may be possible
to impose renormalisation conditions so that the renormalised
composite fields assume the expected continuum behaviour
under parity, up to cutoff effects. Alternatively, one
may be able to identify the parity violating
contributions to the correlation function on the l.h.s. 
of~eq.~(\ref{G_matchII}),
so that the parity conserving QCD correlator $G^\pm_{K\pi}$
can be obtained. Moreover, this is only required in the
asymptotic regime where the matrix element is extracted.

In practice, a combination of both
methods seems advisable. Hence we begin by identifying 
three sources  of parity violating contributions to
the tmQCD correlation function in the l.h.s.~of eq.~(\ref{G_matchII}):
\begin{enumerate}
\item The operator $O_{\rm VA+AV}^\pm$ has linear 
divergences of either parity, cf. eq.~(\ref{Opm_ren}).
\item
In general, the interpolating fields $\tilde \varphi_K, \tilde\varphi_\pi$
may break physical parity. However, 
once the twist angles $\alpha,\beta$ are defined from the renormalised
PCAC and PCVC relations, $\tilde \varphi_K, \tilde \varphi_\pi$ may be
chosen so that parity holds up to O($a$) effects (explicit examples will
be given in the following).
\item
In general, the action will 
have parity breaking O($a$) counterterms. In the
spirit of Symanzik's effective continuum theory, 
these counterterms can be treated as operator
insertions, and thus contribute at O($a$) to 
the correlation function.
\end{enumerate}
As a consequence, unwanted O(1) contributions which originate
from parity violation may arise by combining
the parity violating linear divergence of the four-quark operator
with  parity violating O($a$) terms from either the action 
or the interpolating fields. Hence, if the parity
breaking linear divergence of the four-quark operator 
is subtracted by imposing an appropriate renormalisation condition, 
the parity violating contributions to the correlation function
are O($a$). 

In the following we make two specific choices for the twist
angles  $\alpha$ and $\beta$ (cf.~subsect.~4.1).
For both cases we explain in detail how to obtain the  
$K\to\pi$ matrix elements from the 
tmQCD correlation functions on the l.h.s.~of eq.~(\ref{G_matchII}).

\subsection{The fully twisted case $\alpha=\beta=\pi/2$}

For simplicity we specify local source fields
which, for $\alpha=\beta=\pi/2$ read
\be
  \tilde{\varphi}_K = i(S_\rmR)_{us},\qquad
  \tilde{\varphi}_\pi = (P_\rmR)_{du}.
\label{eq:ptsources}
\ee
We have already pointed out that
the fully twisted case is special in that the
operators of interest are directly matched to each other; thus
the $K^+ \to \pi^+$ correlation function (\ref{eq:GKpi})
simplifies to
\be
G^\pm_{K\pi}(x,y,z) = \langle (P_\rmR)_{du}(x) (O_\rmR)^\pm_{\rm VA+AV}(y)
(S_\rmR)_{us}(z) \rangle_{(\pi/2,\pi/2)} \,\, .
\label{eq:GKpitwisted}
\ee
It follows that, up to cutoff effects, the renormalised four-quark
operators have definite physical parity $\cP$, viz.
\be
   \cP [(O_\rmR)^\pm_{\rm VA+AV}] = +1 \ .
\ee
On the other hand, for the dimension-3 counterterms 
we have 
\be
   \cP[P_{sd}] = -1 \qquad \cP[S_{sd}] = +1 \ ,
\ee
i.e.~the linearly divergent coefficient $c_P^\pm$ 
multiplies a parity odd counterterm, while the finite counterterm 
proportional to $c_S^\pm$ is parity even. 
This implies that, out of the two renormalisation 
conditions required to determine the counterterm
coefficients $c_P^\pm$ and $c_S^\pm$, the one determining
$c_P^\pm$ must restore physical parity of the renormalised operator.
For example, the condition
\be
 \left\langle (O_\rmR)^\pm_{\rm VA+AV}(x) \,\,\, (P_\rmR)_{ds}(y)   
 \right\rangle_{(\pi/2,\pi/2)} = 0,
  \label{eq:parity0}
\ee
enforces parity conservation, as can be seen 
from its translation to standard QCD,
\be
 \left\langle i (O_\rmR)^\pm_{\rm VV+AA}(x) (P_\rmR)_{ds}(y)   
 \right\rangle_{(0,0)}=0.
\ee 
In terms of the bare operators eq.~(\ref{eq:parity0}) reads
\bea
 0 &=& \left\langle O^\pm_{\rm VA+AV}(x) \,\,\, 
   P_{ds}(y)  \right\rangle_{(\pi/2,\pi/2)} 
 \nonumber\\
   &&\mbox{} + c^\pm_P \left\langle P_{sd}(x) \,\,\, P_{ds}(y) 
                      \right\rangle_{(\pi/2,\pi/2)} 
   \nonumber\\
   && \mbox{}+ c^\pm_S \left\langle S_{sd}(x) \,\,\, P_{ds}(y)
                      \right\rangle_{(\pi/2,\pi/2)}.
 \label{eq:rc_parity}
\eea
Note that this equation indeed determines the coefficient $c_P^\pm$, 
as the last term on the r.h.s.~satisfies
\be
  \left\langle S_{sd}(x) \,\,\, 
    P_{ds}(y)\right\rangle_{(\pi/2,\pi/2)}=
  (Z_\rmP Z_\rmS)^{-1}
 \left\langle (S_\rmR)_{sd}(x) \,\,\, (P_\rmR)_{ds}(y)\right\rangle_{(0,0)},
\label{eq:podd}
\ee
and thus vanishes by parity up to O($a$). 
In particular, the prefactor $c^\pm_S$, being of O($1$), does not affect this 
conclusion. 

We may now insert the renormalised operator $(O_\rmR)^\pm_{\rm VA+AV}$
(with $c_P^\pm$ determined as above and $c_S^\pm$ as 
yet undetermined)
in the correlation function~(\ref{eq:GKpitwisted}). 
As the mesonic source operators of eq.(\ref{eq:ptsources}) 
are also parity eigenstates,
\be
  \cP[S_{us}] = -1 \qquad \cP[P_{du}] = -1 \,\, .
\ee
the remaining parity breaking effects in the
correlation function~(\ref{eq:GKpitwisted}) are at most of O($a$), being
combinations of O($a$) effects from either the
action or the mesonic source fields with an O(1) term
of the four-quark operator.
It thus remains to determine the finite counterterm proportional 
to $c_S^\pm$. This can be done
by imposing the condition (\ref{matrix_renorm}) on the
$K^+\to\pi^+$ matrix element which, in the current tmQCD framework,
is obtained from the correlation~(\ref{eq:GKpitwisted}).

\subsection{Sidestepping the power divergence}

As explained in subsection~4.4, the action in the fully twisted
cases can be considered O($a$) improved, provided O($a$) improvement
has been correctly implemented in the massless theory.
In this situation, the above renormalisation procedure 
can be further simplified, as
the determination of the coefficient $c_P^\pm$ from the
condition~(\ref{eq:rc_parity}) can be avoided.

To see this we write explicitly
the correlation $G^\pm_{K\pi}$ in tmQCD:
\bea
G^\pm_{K\pi}(x,y,z) &=& Z_{\rmP} Z_\rmS Z^\pm \bigg [
\langle P_{du}(x) O^\pm_{\rm VA+AV}(y) S_{us}(z) \rangle_{(\pi/2,\pi/2)} 
\nonumber \\
&&\mbox{}+ c_P^\pm \,\, \langle P_{du}(x) P_{sd}(y) S_{us}(z) \rangle_{(\pi/2,\pi/2)} 
\nonumber \\
&&\mbox{}+ c_S^\pm \,\, \langle P_{du}(x) S_{sd}(y) S_{us}(z) \rangle_{(\pi/2,\pi/2)} 
\bigg ]
\label{eq:G-qcd-tmqcd}
\eea
and examine in detail the correlation proportional to $c_P^\pm$ on the r.h.s..
The operator product $P_{du} P_{sd} S_{us}$ is an odd parity eigenstate, giving
vanishing O(1) contributions. The O($a$) corrections (which 
give rise to O(1) contributions to the original correlation function
$G^\pm_{K\pi}$) may arise from three sources, as
listed in subsect.~\ref{subsec:corrls}:

\begin{enumerate}
\item The parity-even O($a$) correction to the operator insertion 
$P_{sd}$ has the form $a(\mu_s+\mu_d)S_{sd}$, due to the symmetries of the 
lattice action. This term amounts to a re-definition of $c_S^\pm$
and can thus be ignored.

\item The even parity O($a$) corrections to the kaon source $S_{us}$
are proportional to $(\mu_u+\mu_s)P_{us}$, while the analogous
counterterm for the pion only appears at O($a^2$) 
(for mass degenerate light quarks);
the O($a$) correction to the Fourier transformed correlation function
is then given by
\be
   a^6\sum_{\bfx,\bfz}\rme^{i(\bfp\bfx-\bfq\bfz)}\langle P_{du}(x)
    P_{sd}(0) P_{us}(z) \rangle_{(\pi/2,\pi/2)} 
   \;\buildrel{z_0\rightarrow-\infty}\over{\propto}\;
 \rme^{z_0 E_{K^S}(\bfq)},
\ee
where we have exhibited the expected asymptotic
factor for large negative $z_0$, and $K^S$ denotes the
lowest scalar excitations in the kaon channel.
It is not clear what the exact nature of this state is.
Candidates are one-particle states with the opposite parity, or
multi-particle states. For instance,
a $J=0$ two-particle state consisting of a kaon and a neutral
pion shares all the lattice symmetries with the kaon,
and might therefore be the next lightest state with opposite physical parity.
In any case, we expect the energies of the excited states to
be much higher than the kaon energies themselves.
Therefore, in the asymptotic regime, the relative suppression factor
$\exp\{z_0\left[E_{K^S}(\bfq)-E_K(\bfq)\right]\}$ of the second term
on the rhs of eq.~(\ref{eq:G-qcd-tmqcd}) should be significant.

\item Finally, as the O($a$) contribution of the action vanishes
with the appropriate choice of the counterterm basis (cf.~subsect.~4.4), 
the action contributes at most to O($a^2$).
This combines with the linear divergence of the operator
to yield an O($a$) effect to $G^\pm_{K\pi}$.
Here it is crucial that the action be O($a$) improved, 
as the insertion of action
counterterms into the correlation function comes with a 
space-time integration, which makes the effect non-local.
In particular, it would not be possible to argue in terms of the 
asymptotic regime of the correlation function 
as we did in (ii) with the localised counterterms to the kaon source field.
\end{enumerate}

We conclude that in the fully twisted case the $c_P^\pm$ term in
eq.~(\ref{eq:G-qcd-tmqcd}) need
not be subtracted explicitly, as it leads at most to O($1$) contributions
to $G^\pm_{K\pi}$ which are exponentially suppressed, due to the 
energy gap between scalar and pseudoscalar sector.
Remarkably, the only subtraction to be carried out explicitly is
the finite, parity conserving counterterm proportional to $c_S^\pm$.
One is thus left with a renormalisation structure that closely
resembles the one found in regularisations that preserve chiral symmetry.

\subsection{The partially twisted case with $\beta=0$}

In this case, the operator $O^\pm_{\rm VA+AV}$ is mapped 
to a linear combination of parity odd and parity even 
pieces~(cf.~\req{Opm_twist_2}).
Therefore, it is not clear how to impose a parity restoration
condition for the operator. We therefore proceed in two steps:
first we subtract the two power divergent counterterms
with some arbitrary prescription, e.g.~by imposing the conditions
\bea 
 \left\langle (\bar{O})^\pm_{\rm VA+AV}(x) (P)_{ds}(y)   
 \right\rangle_{(\pi/2,0)} &=& 0,
  \label{eq:intermediate_1}\\
 \left\langle (\bar{O})^\pm_{\rm VA+AV}(x) (S)_{ds}(y)   
 \right\rangle_{(\pi/2,0)} &=& 0.
  \label{eq:intermediate_2}
\eea
Here $\bar{O}^\pm_{\rm VA+AV}$ is the subtracted bare
operator,
\be
   \bar{O}^\pm_{\rm VA+AV} =   {O}^\pm_{\rm VA+AV} 
   +\bar{c}_P^\pm P_{sd} + \bar{c}_S^\pm  S_{sd}
\ee
and the coefficients $\bar{c}^\pm_{P,S}$ are determined by 
eqs.~(\ref{eq:intermediate_1},\ref{eq:intermediate_2}).
This bare operator is only logarithmically divergent, and 
its renormalised counterpart is obtained by multiplicative
rescaling with the usual renormalisation constant,
\be
    (\bar{O}_\rmR)^\pm_{\rm VA+AV} =
     Z^\pm \bar{O}^\pm_{\rm VA+AV}.
\ee 
However, such a renormalised operator does not have 
the physical parity properties that ensure the restoration of parity
in the continuum limit.
Rather, a correctly renormalised operator will be of the form
\be
   {(O_\rmR)}^\pm_{\rm VA+AV} = 
   Z^\pm \left\{\bar{O}^\pm_{\rm VA+AV} 
   +\Delta{c}_P^\pm P_{sd} + \Delta{c}_S^\pm  S_{sd}\right\},   
\ee 
with still unknown, but {\em finite} coefficients $\Delta{c}^\pm_{P,S}$.
These should be determined so
that the hadronic matrix element of the renormalised operator
does not contain unwanted contributions of the wrong physical parity.
This does not necessarily determine the renormalised operator itself. 
Indeed, it turns out that only a single 
coefficient needs to be determined, the other giving 
rise to contributions which are either
of O($a$) or exponentially suppressed or both, depending
on the choice of interpolating kaon and pion fields.

To see this more explicitly, we have a closer look at the 
additive counterterms, $P_{sd}$ and $S_{sd}$. 
It is convenient to split both fields in 
pieces which are either even or odd under 
the physical parity transformation.
\be
    P_{ds}=\left(P_{ds}\right)_{\rm even} + \left(P_{ds}\right)_{\rm odd},
   \qquad
    S_{ds}=\left(S_{ds}\right)_{\rm even} + \left(S_{ds}\right)_{\rm odd}.
\ee
From the correspondence between the renormalised fields in tmQCD
and standard QCD (cf.~subsection~3.2) one then infers 
the explicit expressions,
\bea
  \left(P_{ds}\right)_{\rm even} &=& \sin^2\left(\dfrac{\alpha}{2}\right)P_{ds}
   + Z_\rmS/Z_\rmP \dfrac{i}{2} \sin(\alpha)S_{ds},\label{pdseven}\\
  \left(P_{ds}\right)_{\rm odd} &=& \cos^2\left(\dfrac{\alpha}{2}\right)P_{ds}
   - Z_\rmS/Z_\rmP \dfrac{i}{2} \sin(\alpha)S_{ds},\\
  \left(S_{ds}\right)_{\rm even} &=& \cos^2\left(\dfrac{\alpha}{2}\right)S_{ds}
   - Z_\rmP/Z_\rmS \dfrac{i}{2} \sin(\alpha)P_{ds},\label{sdseven}\\
  \left(S_{ds}\right)_{\rm odd} &=& \sin^2\left(\dfrac{\alpha}{2}\right)S_{ds}
   + Z_\rmP/Z_\rmS \dfrac{i}{2} \sin(\alpha)P_{ds},
\eea
where terms of O($a$) have been neglected.
To discuss contributions to the correlation function from which
$K^+\to\pi^+$ matrix elements are extracted we 
first assume that the pion and kaon source fields have been
chosen with definite physical parity, i.e.~up to cutoff effects
they generate excitations with only pion or kaon quantum numbers and
thus of odd physical parity. It is then clear that the odd parity
parts of the renormalised operator can only generate O($a$) contributions
to the correlation function. Neglecting these
we are thus left with contributions of the parity even counterterms,
i.e.~only the combination,
\be
  \Delta{c}_P^\pm \left(P_{sd}\right)_{\rm even} 
+ \Delta{c}_S^\pm \left(S_{sd}\right)_{\rm even},   
\ee
contributes at O(1). Moreover, as can be inferred from
eqs.~(\ref{pdseven},\ref{sdseven}), the parity even operators
are proportional to each other 
\be
     \left(P_{ds}\right)_{\rm even} \propto
      \left(S_{ds}\right)_{\rm even},
\ee
up to terms of O($a$).
This means that there is only a linear combination of
counterterms which contribute at O(1), which can be
written as
\be
   \left\{\Delta{c}_P^\pm  
          - Z_\rmP/Z_\rmS \dfrac{2i\cos^2(\alpha/2)}{\sin(\alpha)} 
           \Delta{c}_S^\pm
  \right\} \left(P_{sd}\right)_{\rm even}
   \equiv   d_P^{\pm} \left(P_{sd}\right)_{\rm even}.
\ee
Moreover, it is not necessary to determine the parity even part
of $P_{sd}$, as the corresponding parity odd part does 
only contribute at O($a$). It is
therefore enough to insert for the renormalised operator
\be
   {(O_\rmR)}^\pm_{\rm VA+AV} \longrightarrow 
   Z^\pm \left\{\bar{O}^\pm_{\rm VA+AV} +{d}_P^\pm P_{sd} 
                     \right\}.
  \label{replacement}
\ee
In other words: for insertions into the $K\to\pi$ correlation 
function with meson sources of definite parity, the
replacement (\ref{replacement}) yields equivalent results
up to O($a$) effects. The remaining coefficient $d^\pm_{P}$
is then determined non-perturbatively by the condition (\ref{matrix_renorm})
on the hadronic matrix element.

Finally, by the same arguments as in the preceding subsection
we may relax the requirement that the meson sources
have a definite parity. This is again due to the
expected energy gap between scalar and pseudoscalar channels,
which implies an exponential suppression of the wrong
parity contributions. The sources
with a definite parity therefore just provide an additional
suppression by a power of $a$ of these terms, which is
not necessary for the extraction of the matrix element.

In conclusion, in the partially twisted case with $\beta=0$, 
the two linearly divergent counterterms can be subtracted
non-perturbatively by imposing two independent but
otherwise arbitrary conditions on correlation functions
involving the four-quark operator.
Once this has been achieved it turns out that
wrong parity contributions to the correlation function
are of O($a$) provided that the mesonic source fields have odd parity.
Otherwise they are in any case exponentially suppressed in the asymptotic
regime where the matrix element is extracted. A remaining
finite subtraction constant is determined in the usual
way by the matrix element itself from eq.~(\ref{matrix_renorm}).

\subsection{$K^0\to\pi^0$ matrix elements}

As stated earlier, the breaking of isospin symmetry in twisted mass
lattice QCD obscures the decomposition of four-quark operators into
$\Delta I=1/2$ and $\Delta I=3/2$ pieces. 
In order to obtain both effective couplings, $g_8$ and $g_{27}$,
one may therefore opt to compute, 
besides the $K^+\to\pi^+$ transition, also $K^0\to\pi^0$ matrix elements.

The strategy used for $K^+\to\pi^+$ matrix elements can be
taken over virtually unchanged. However, a
technical complication is generated by the fact that
disconnected diagrams are now possible, as the self-contraction
of the pion source does not vanish by flavour symmetry.
Moreover, one may be worried about new power divergences,
which may arise in the pion source field.
For instance, the field
\be
    \tilde{\varphi}_{\pi^0} = \bar{u}\gamma_5 u-\bar{d}\gamma_5d,
\ee
suffers from a quadratic divergence $\propto \mu_l/a^2$.
However, we note that this can be easily bypassed 
with the choice of the axial current,
\be
    \tilde{\varphi}_{\pi^0} = A_0^3,
\ee
which, due to the O($4$) vector structure, does not suffer 
from additive renormalisation.
Hence we conclude that, beyond possible practical problems
relating to the quality of the signal in the numerical
simulations, the $K^0\to\pi^0$ matrix elements do not present
any further theoretical challenge.

\subsection{A comment on $K\to\pi\pi$ transitions}

In principle, there is no obstacle to applying the strategies 
described here to the physical $K\to\pi\pi$ amplitudes. 
Parity is changed in this transition, so that only the 
parity odd operator contributes to the matrix element.
This operator has nice renormalisation properties
already with standard Wilson quarks (\ref{VAAV_renW}), so that 
the potential gain by using twisted mass QCD mainly 
consists in the elimination of unphysical zero modes.

A direct mapping of the operator $O^\pm_{\rm VA+AV}$
in the twisted basis to the parity odd operator
in the standard basis is obtained for
twist angles $\alpha=-\beta=\pi/2$ [cf.~eq.~(\ref{Kpipi})].
Compared to the case of $\beta=\pi/2$ this implies that
the sign of $\mu_s$ and $\mu_c$ are reversed. Alternatively,
one may again set $\beta=0$, and project on the
desired matrix element by choosing interpolating fields
for the two-pion and kaon states with the correct physical
parity.  However, a practical problem may arise from 
isospin symmetry breaking in lattice regularised tmQCD:
for instance, the lattice symmetries do not distinguish 
the $I=0$ two-pion state from the state with a single neutral pion.
As the energy of a single pion is lower than the energy of 
the two-pion state,  one needs to carry out a multistate analysis which
renders this approach more complicated. 
Nevertheless, the good news remains that unphysical zero modes
can be eliminated whilst the renormalisation properties
are no worse than in standard lattice QCD with Wilson type quarks.

\section{Conclusions}

Twisted mass lattice QCD provides a viable framework for 
a determination of $g_8$ and $g_{27}$, 
by allowing for comparatively cheap numerical computations 
of $K\to\pi$ matrix elements at light meson masses.
In order to facilitate future practical implementations of this 
strategy, we summarise our results in the form of a recipe:
\begin{enumerate}
\item
Choose the number of dynamical quarks to be 0, 2,3 or 4,
with mass degenerate up and down quarks;
the recommended twist angles are then $(\alpha,\beta)=(\pi/2,\pi/2)$
or $(\alpha,\beta)=(\pi/2,0)$. The latter choice avoids 
a complex quark determinant if more than 2 quarks are dynamical.
\item
Translate the twist angles to bare mass parameters:
$(\alpha,\beta)=(\pi/2,\pi/2)$ means that all standard mass parameters
are tuned to the critical mass $m_{\rm cr}$, which can be obtained
from the PCAC relation as usual. The physical quark masses are
then determined by the twisted mass parameters $\mu_l,\mu_s,\mu_c$
which are fixed by matching appropriate experimental quantities.

The second option, $(\pi/2,0)$, corresponds
to $\mu_s=\mu_c=0$. For given mass parameters $m_s, m_c$, one
then needs to determine $m_l$ so that the axial current 
in the light quark sector is conserved. A simplification occurs
with quenched strange and charm quarks, where this is equivalent to
setting $m_l=m_{\rm cr}$, independently of $m_s,m_c$.
The physical quark masses are determined in terms of $\mu_l, m_s$ and $m_c$.

\item
Choose your preferred mesonic source fields for the kaon and pion;
these need not be of definite physical parity, however, they should 
have a sufficiently large overlap with the physical kaon or pion state.
In the case of full twist, $(\pi/2,\pi/2)$, there are two ways to
proceed:
\begin{itemize}
  \item if O($a$) improvement of the massless action and the axial current
   is implemented via the counterterms $\propto c_{\rm sw}$ and 
   $\propto c_{\rm A}$ respectively, 
   the tmQCD action can be considered O($a$) improved and
   {\em the linear divergence in the four-quark operator need not be
   subtracted}. Just compute the bare 3-point function 
   and the pion and kaon propagators, and
   go to the asymptotic regime to obtain the matrix element
   up to O($a$) effects. Requiring the matrix element to vanish
   at zero meson momenta fixes the coefficient $c_S^\pm$ of the 
   finite counterterm.
  \item if O($a$) improvement is not implemented, one needs
   to first determine $c_P^\pm$  e.g.~by imposing the
   parity restoration condition (\ref{eq:parity0}). Using the 
   resulting subtracted operator one may proceed exactly as above.
\end{itemize}
In the case of partial twist, $(\pi/2,0)$, one
first subtracts the linear divergences by imposing two arbitrary
renormalisation conditions 
such as~(\ref{eq:intermediate_1},\ref{eq:intermediate_2}). 
Then one proceeds as above, using the subtracted operator. 
To fix the finite counterterm
through the $K\to\pi$ matrix element at vanishing meson momenta, one
may use the substitution~(\ref{replacement}) 
which avoids determining the parity even part of the counterterm.
\item
Repeat the calculation for a sequence of lattice spacings,
keeping the physical scales fixed, and extrapolate 
to the continuum limit. Note that there exist
various alternative strategies to compute
the same matrix element. By varying the choice of twist angles, 
the way in which the linear divergences are subtracted, or
the  choice of mesonic source fields, one should obtain
always the same matrix element up to O($a$) effects.
Thus, one may hope to improve the 
control of the continuum limit by pursuing 
a couple of alternative strategies in parallel.
\item
The continuum results obtained for various meson momenta
and energies should be well described by leading order 
chiral perturbation theory, provided the meson masses
and momenta are small enough. If so, the $K^+\to\pi^+$
matrix element allows to determine 
the linear combination $\frac23 g_{27}+g_8$, while the
matrix element for neutral mesons yields $g_{27}-g_8$
(cf.~subsect.~2.3). If the $\Delta I=1/2$ rule can be 
explained within QCD, one would expect 
the lattice results to reproduce the hierarchy 
$g_8\gg g_{27}$ obtained for the phenomenological estimates~(\ref{g_pheno}).
\end{enumerate}
The strategy described above offers major improvements
over previous attempts to compute $K\to\pi$ transitions
with Wilson-type quarks. Nevertheless,
it is conceivable that further theoretical developments, 
along the lines of refs.~\cite{FR,Frezzotti:2003xj}, may lead 
to interesting variations and/or alternatives to the current work.

\subsection*{Acknowledgments}
We thank H.~Wittig for a critical reading
of the manuscript, and A.~Donini, R.~Frezzotti and G.C.~Rossi
for discussions. 
The hospitality of the Theory Division of CERN (C.P.),
the Physics Department at the University of Rome ``Tor Vergata'' (C.P.),
and the IFT at the Autonomous University of Madrid (C.P., A.V.)
is gratefully acknowledged.
C.P.~has been partially supported 
through the European Community's Human Potential Programme 
under contract HPRN-CT-2000-00145, Hadrons/Lattice QCD.


\begin{thebibliography}{99}

% section 1

\bibitem{analytic}
W.A.~Bardeen, A.J.~Buras and J.M.~Gerard, Phys. Lett. B {\bf 192} (1987) 138;\\
S.~Bertolini, J.O.~Eeg, M.~Fabbrichesi and E.I.~Lashin, 
Nucl. Phys. B {\bf 514} (1998) 63, and references therein;\\
J.~Bijnens and J.~Prades, JHEP {\bf 9901} (1999) 023;\\
%[arXiv:hep-ph/9811472].
%%CITATION = HEP-PH 9811472;%%
T.~Hambye, S.~Peris and E.~de~Rafael, JHEP {\bf 0305} (2003) 027.
%[arXiv:hep-ph/0305104].
%%CITATION = HEP-PH 0305104;%%

%\cite{wil}
\bibitem{wil}
D.~Be\'cirevi\'c et al. [SPQ$_{\rm CD}$R Collaboration],
Nucl.\ Phys.\ Proc.\ Suppl.\  {\bf 119} (2003) 359.
%[arXiv:hep-lat/0209136].
%%CITATION = NUPHZ,119,359;%%

%\cite{stag}
\bibitem{stag}
D.~Pekurovski and G.~Kilcup, 
Phys.\ Rev.\ D {\bf 64} (2001) 074502;\\
%[arXiv:hep-lat/9812019].
%%CITATION = PHRVA,D64,074502;%%
T.~Bhattacharya et al.,
Nucl.\ Phys.\ Proc.\ Suppl.\  {\bf 129} (2004) 257.
%[arXiv:hep-lat/0309105].
%%CITATION = NUPHZ,129,257;%%

%\cite{GW}
\bibitem{GW}
T.~Blum et al. [RBC Collaboration],
Phys.\ Rev.\ D {\bf 68} (2003) 114506;\\
%[arXiv:hep-lat/0110075].
%%CITATION = PHRVA,D68,114506;%%
A.~Soni, Pramana {\bf 62} (2004) 415;\\
J.I.~Noaki et al. [CP-PACS Collaboration],
Phys.\ Rev.\ D {\bf 68} (2003) 014501.
%[arXiv:hep-lat/0108013].
%%CITATION = PHRVA,D68,014501;%%

%\cite{becir}
\bibitem{becir}
D.~Be\'cirevi\'c, 
Nucl.\ Phys.\ Proc.\ Suppl.\  {\bf 129} (2004) 34.
%%CITATION = NUPHZ,129,034;%%

%\cite{MaianiTesta}
\bibitem{MaianiTesta}
L.~Maiani and M.~Testa
Phys.\ Lett.\ B {\bf 245} (1990) 585.

%\cite{Lellouch:2000pv}
\bibitem{Lellouch:2000pv}
L.~Lellouch and M.~L\"uscher,
%``Weak transition matrix elements from finite-volume correlation  functions,''
Commun.\ Math.\ Phys.\  {\bf 219} (2001) 31.
%[arXiv:hep-lat/0003023].
%%CITATION = HEP-LAT 0003023;%%

%\cite{Lin:2001ek}
\bibitem{Lin:2001ek}
C.~J.~Lin, G.~Martinelli, C.~T.~Sachrajda and M.~Testa,
%``K $\to$ pi pi decays in a finite volume,''
Nucl.\ Phys.\ B {\bf 619} (2001) 467.
%[arXiv:hep-lat/0104006].
%%CITATION = HEP-LAT 0104006;%%

%\cite{KimChrist}
\bibitem{KimChrist}
C.~Kim and N.H.~Christ, Nucl.\ Phys.\ Proc.\ Suppl.\  {\bf 119} (2003) 365.
%%CITATION = NUPHZ,119,365;%%

%\cite{Bernard:wf}
\bibitem{Bernard:wf}
C.~W.~Bernard, T.~Draper, A.~Soni, H.~D.~Politzer and M.~B.~Wise,
%C.~W.~Bernard et al.,
%``Application Of Chiral Perturbation Theory To K $\to$ 2 Pi Decays,''
Phys.\ Rev.\ D {\bf 32} (1985) 2343.
%%CITATION = PHRVA,D32,2343;%%

%\cite{Bernard:1987pr}
\bibitem{Bernard:1987pr}
C.~W.~Bernard, T.~Draper, G.~Hockney and A.~Soni,
%``Recent Developments In Weak Matrix Element Calculations,''
Nucl.\ Phys.\ Proc.\ Suppl.\  {\bf 4} (1988) 483.
%%CITATION = NUPHZ,4,483;%%

%\cite{Dawson:1997ic}
\bibitem{dih_review_2}
C.~Dawson et al., 
%C.~Dawson, G.~Martinelli, G.~C.~Rossi, C.~T.~Sachrajda, S.~R.~Sharpe, 
%M.~Talevi and M.~Testa,
%``New lattice approaches to the Delta(I) = 1/2 rule,''
Nucl.\ Phys.\ B {\bf 514} (1998) 313.
%[arXiv:hep-lat/9707009].
%%CITATION = HEP-LAT 9707009;%%

%\cite{Laiho:2002jq}
\bibitem{Laiho:2002jq}
J.~Laiho and A.~Soni,
%``On lattice extraction of K $\to$ pi pi amplitudes to O(p**4) 
% in chiral  perturbation theory,''
Phys.\ Rev.\ D {\bf 65} (2002) 114020;
%[arXiv:hep-ph/0203106];
%%CITATION = HEP-PH 0203106;%%
%\cite{Laiho:2003uy}
%\bibitem{Laiho:2003uy}
%J.~Laiho and A.~Soni,
%``Lattice extraction of K $\to$ pi pi amplitudes to NLO in 
% partially  quenched and in full chiral perturbation theory,''
hep-lat/0306035.
%%CITATION = HEP-LAT 0306035;%%

%\cite{Lin:2002nq}
\bibitem{Lin:2002nq}
C.~J.~Lin, G.~Martinelli, E.~Pallante, C.~T.~Sachrajda and G.~Villadoro,
%C.~J.~Lin et al.,
%``K+ $\to$ pi+ pi0 decays on finite volumes and 
% at next-to-leading order in the chiral expansion,''
Nucl.\ Phys.\ B {\bf 650} (2003) 301.
%[arXiv:hep-lat/0208007].
%%CITATION = HEP-LAT 0208007;%%


%\cite{Capitani:2000da}
\bibitem{Capitani:2000da}
S.~Capitani and L.~Giusti,
%``Perturbative renormalization of 
% weak-Hamiltonian four-fermion operators  with
%overlap fermions,''
Phys.\ Rev.\ D {\bf 62} (2000) 114506.
%[arXiv:hep-lat/0007011].
%%CITATION = HEP-LAT 0007011;%%

%\cite{Capitani:2000bm}
\bibitem{GW_theor}
S.~Capitani and L.~Giusti,
%``Analysis of the Delta(I) = 1/2 rule and epsilon'/epsilon with overlap
%fermions,''
Phys.\ Rev.\ D {\bf 64} (2001) 014506.
%[arXiv:hep-lat/0011070].
%%CITATION = HEP-LAT 0011070;%%

%\cite{Aoki:2000ee}
\bibitem{Aoki:2000ee}
S.~Aoki and Y.~Kuramashi,
%``Perturbative renormalization factors of Delta(S) = 1 four-quark  operators
%for domain-wall QCD,''
Phys.\ Rev.\ D {\bf 63} (2001) 054504.
%[arXiv:hep-lat/0007024].
%%CITATION = HEP-LAT 0007024;%%

%\cite{Bochicchio:1985xa}
\bibitem{Bochicchio:1985xa}
M.~Bochicchio, L.~Maiani, G.~Martinelli, G.~C.~Rossi and M.~Testa,
%M.~Bochicchio et al.,
%``Chiral Symmetry On The Lattice With Wilson Fermions,''
Nucl.\ Phys.\ B {\bf 262} (1985) 331.
%%CITATION = NUPHA,B262,331;%%

%\cite{Maiani:1986db}
\bibitem{Maiani:1986db}
L.~Maiani, G.~Martinelli, G.~C.~Rossi and M.~Testa,
%``The Octet Nonleptonic Hamiltonian And Current Algebra On The Lattice With
%Wilson Fermions,''
Nucl.\ Phys.\ B {\bf 289} (1987) 505.
%%CITATION = NUPHA,B289,505;%%

%\cite{Hernandez:2002ds}
\bibitem{Hernandez:2002ds}
P.~Hern\'andez and M.~Laine,
%``Correlators of left charges and weak operators in finite volume chiral
%perturbation theory,''
JHEP {\bf 0301} (2003) 063.
%[arXiv:hep-lat/0212014].
%%CITATION = HEP-LAT 0212014;%%

%\cite{Giusti:2004an}
\bibitem{Giusti:2004an}
L.~Giusti, P.~Hern\'andez, M.~Laine, P.~Weisz and H.~Wittig,
%``A strategy to study the role of the charm quark in explaining the Delta(I) =
%1/2 rule,''
hep-lat/0407007.
%%CITATION = HEP-LAT 0407007;%%

%\cite{Pena:2002wj}
\bibitem{dih_proc}
C.~Pena, S.~Sint and A.~Vladikas,
%``Twisted mass QCD and the Delta(I) = 1/2 rule,''
Nucl.\ Phys.\ Proc.\ Suppl.\  {\bf 119} (2003) 368;
%[arXiv:hep-lat/0209045].
%%CITATION = HEP-LAT 0209045;%%
%\cite{Pena:2003pa}
%\bibitem{Pena:2003pa}
%C.~Pena, S.~Sint and A.~Vladikas,
%``Towards a determination of g(8) and g(27) from twisted mass lattice  QCD,''
Nucl.\ Phys.\ Proc.\ Suppl.\  {\bf 129} (2004) 263.
%%CITATION = NUPHZ,129,263;%%


% phenomenological estimates of g_8 and g_27

\bibitem{lo}
A.~Pich, B.~Guberina and E.~de Rafael,
%``Problem With The Delta I = 1/2 Rule In The Standard Model,''
Nucl.\ Phys.\ B {\bf 277} (1986) 197;
%%CITATION = NUPHA,B277,197;%%
%
A.~Pich and E.~de Rafael,
%``Four Quark Operators And Nonleptonic Weak Transitions,''
Nucl.\ Phys.\ B {\bf 358} (1991) 311.
%%CITATION = NUPHA,B358,311;%%

%\cite{Gasser:1983yg}
\bibitem{Gasser:1983yg}
J.~Gasser and H.~Leutwyler,
%``Chiral Perturbation Theory To One Loop,''
Annals Phys.\  {\bf 158} (1984) 142;
%%CITATION = APNYA,158,142;%%
%\cite{Gasser:1984gg}
%\bibitem{Gasser:1984gg}
%J.~Gasser and H.~Leutwyler,
%``Chiral Perturbation Theory: Expansions In The Mass Of The Strange Quark,''
Nucl.\ Phys.\ B {\bf 250} (1985) 465.
%%CITATION = NUPHA,B250,465;%%

% section 2

%     no contribution from $g_8'$:
\bibitem{rc}
R.J.~Crewther,
%``Chiral Reduction Of K $\to$ 2 Pi Amplitudes,''
Nucl.\ Phys.\ B {\bf 264} (1986) 277.
%%CITATION = NUPHA,B264,277;%%

%% phases
\bibitem{gm}
J.~Gasser and U.G.~Meissner,
%``On The Phase Of Epsilon-Prime,''
Phys.\ Lett.\ B {\bf 258} (1991) 219.
%%CITATION = PHLTA,B258,219;%%

%\cite{Itzykson:rh}
\bibitem{Itzykson}
C.~Itzykson and J.~B.~Zuber,
``Quantum Field Theory,'' McGraw-Hill (1980), New York, U.S.A.

% difficulties for K ->pi amplitudes with Wilson quarks
%\cite{Gavela:ws}
\bibitem{Gavela:ws}
M.~B.~Gavela, L.~Maiani, S.~Petrarca, G.~Martinelli and O.~P\`ene,
%M.~B.~Gavela et al.,
%``First Results For The Delta I = 1/2 Amplitude 
%In K Decays, With Quenched Lattice QCD And Wilson Fermions,''
Phys.\ Lett.\ B {\bf 211} (1988) 139.
%%CITATION = PHLTA,B211,139;%%

%\cite{Donini:1999sf}
\bibitem{Donini:1999sf}
A.~Donini, V.~Gim\'enez, G.~Martinelli, M.~Talevi and A.~Vladikas,
%A.~Donini et al.,
%``Non-perturbative renormalization of lattice four-fermion operators  
% without power subtractions,''
Eur.\ Phys.\ J.\ C {\bf 10} (1999) 121.
%[arXiv:hep-lat/9902030].
%%CITATION = HEP-LAT 9902030;%%

%\cite{Frezzotti:1999vv}
\bibitem{tmQCD_proc1}
R.~Frezzotti, P.~A.~Grassi, S.~Sint and P.~Weisz,
%``A local formulation of lattice QCD without unphysical fermion zero  modes,''
Nucl.\ Phys.\ Proc.\ Suppl.\  {\bf 83} (2000) 941.
%[arXiv:hep-lat/9909003].
%%CITATION = HEP-LAT 9909003;%%

%\cite{Frezzotti:2000nk}
\bibitem{tmQCD_1}
R.~Frezzotti, P.~A.~Grassi, S.~Sint and P.~Weisz [ALPHA Collaboration],
%``Lattice QCD with a chirally twisted mass term,''
JHEP {\bf 0108} (2001) 058.
%[arXiv:hep-lat/0101001].
%%CITATION = HEP-LAT 0101001;%%

%\cite{Frezzotti:2001du}
\bibitem{Frezzotti:2001du}
R.~Frezzotti and S.~Sint,
%``Some remarks on O(a) improved twisted mass QCD,''
Nucl.\ Phys.\ Proc.\ Suppl.\  {\bf 106} (2002) 814.
%[arXiv:hep-lat/0110140].
%%CITATION = HEP-LAT 0110140;%%

%\cite{Luscher:1996sc}
\bibitem{Luscher:1996sc}
M.~L\"uscher, S.~Sint, R.~Sommer and P.~Weisz,
%``Chiral symmetry and O(a) improvement in lattice QCD,''
Nucl.\ Phys.\ B {\bf 478} (1996) 365.
%[arXiv:hep-lat/9605038].
%%CITATION = HEP-LAT 9605038;%%


% no problem with determinant sign fluctuations:

%\cite{Kaneko:2003re}
\bibitem{Aoki:2003re}
T.~Kaneko et al. [CP-PACS Collaboration],
%``Light hadron spectrum in three-flavor QCD 
% with O(alpha)-improved Wilson quark
%action,''
Nucl.\ Phys.\ Proc.\ Suppl.\  {\bf 129} (2004) 188.
%[arXiv:hep-lat/0309137].
%%CITATION = HEP-LAT 0309137;%%


%\cite{Frezzotti:2003xj}
\bibitem{Frezzotti:2003xj}
R.~Frezzotti and G.~C.~Rossi,
%``Twisted-mass lattice QCD with mass non-degenerate quarks,''
Nucl.\ Phys.\ Proc.\ Suppl.\  {\bf 128} (2004) 193.
%[arXiv:hep-lat/0311008].
%%CITATION = HEP-LAT 0311008;%%

%\cite{Bhattacharya:1999yg}
\bibitem{Bhattacharya:1999yg}
T.~Bhattacharya, R.~Gupta, W.~J.~Lee and S.~R.~Sharpe,
%``Non-perturbative improvement of bilinears in unquenched QCD,''
Nucl.\ Phys.\ Proc.\ Suppl.\  {\bf 83} (2000) 902.
%[arXiv:hep-lat/9909092].
%%CITATION = HEP-LAT 9909092;%%

%\cite{Bhattacharya:2003nd}
\bibitem{Bhattacharya:2003nd}
T.~Bhattacharya, R.~Gupta, W.~J.~Lee, S.~R.~Sharpe and J.~M.~S.~Wu,
%T.~Bhattacharya et al.,
%``Improved bilinears in unquenched lattice QCD,''
Nucl.\ Phys.\ Proc.\ Suppl.\  {\bf 129} (2004) 441.
%[arXiv:hep-lat/0309087].
%%CITATION = HEP-LAT 0309087;%%

%\cite{tmQCDimprove}
\bibitem{tmQCDimprove}
R.~Frezzotti, S.~Sint and P.~Weisz  [ALPHA Collaboration],
%``O(a) improved twisted mass lattice QCD,''
JHEP {\bf 0107} (2001) 048.
%[arXiv:hep-lat/0104014].
%%CITATION = HEP-LAT 0104014;%%

% non-perturbative renormalization techniques

%\cite{Martinelli:1994ty}
\bibitem{Martinelli:1994ty}
G.~Martinelli, C.~Pittori, C.~T.~Sachrajda, M.~Testa and A.~Vladikas,
%G.~Martinelli et al.,
%``A General method for nonperturbative renormalization of lattice operators,''
Nucl.\ Phys.\ B {\bf 445} (1995) 81.
%[arXiv:hep-lat/9411010].
%%CITATION = HEP-LAT 9411010;%%

%\cite{Jansen:1995ck}
\bibitem{Jansen:1995ck}
K.~Jansen {\it et al.},
%``Non-perturbative renormalization of lattice QCD at all scales,''
Phys.\ Lett.\ B {\bf 372} (1996) 275.
%[arXiv:hep-lat/9512009].
%%CITATION = HEP-LAT 9512009;%%

%\cite{PenaBK}
\bibitem{PenaBK}
%\cite{Guagnelli:xz}
%M.~Guagnelli et al. [ALPHA Collaboration],
M.~Guagnelli, J.~Heitger, C.~Pena, S.~Sint and A.~Vladikas [ALPHA Collaboration],
%``K0 Anti-K0 Mixing From The Schroedinger Functional And Twisted Mass  QCD,''
Nucl.\ Phys.\ Proc.\ Suppl.\  {\bf 106} (2002) 320;
%[arXiv:hep-lat/0110097].
%%CITATION = HEP-LAT 0110097;
%``Non-perturbative scale evolution of four-fermion operators,''
Nucl.\ Phys.\ Proc.\ Suppl.\  {\bf 119} (2003) 436.
%[arXiv:hep-lat/0209046].
%%CITATION = HEP-LAT 0209046;%%


\bibitem{FR}
R.~Frezzotti and G.~C.~Rossi,
%``Chirally improving Wilson fermions. I: O(a) improvement,''
JHEP {\bf 0408} (2004) 007;
%[arXiv:hep-lat/0306014].
%%CITATION = HEP-LAT 0306014;%%
hep-lat/0407002.
%%CITATION = HEP-LAT 0407002;%%

\end{thebibliography}
\end{document}